\documentclass[sigconf]{acmart}

\usepackage{booktabs} 
\usepackage{color, colortbl}
\usepackage{booktabs} 
\usepackage{graphicx}
\usepackage{courier}
\usepackage{makecell}
\usepackage{float}
\usepackage{bm}
\usepackage{multirow}
\usepackage[flushleft]{threeparttable}
\usepackage{flushend}
\usepackage{bbm}
\usepackage{mathrsfs}
\usepackage{enumitem}
\usepackage{subfigure}

\usepackage[ruled, linesnumbered]{algorithm2e} 
\newcommand{\ignore}[1]{}

\newcommand{\etal}{\emph{et al. }}
\usepackage{etoolbox}

\setcopyright{none}
\definecolor{xlcolor}{RGB}{218,165,32}

\title{Pre-trained Language Model for Web-scale Retrieval in Baidu Search}

\author{Yiding Liu$^\#$, Guan Huang$^\#$, Jiaxiang Liu, Weixue Lu, Suqi Cheng, Yukun, Li, Daiting Shi, Shuaiqiang Wang, Zhicong Cheng, Dawei Yin$^*$}

\affiliation{
  \institution{Baidu Inc., China}
}
\email{{liuyiding.tanh,chengsuqi}@gmail.com}
\email{{huangguan01,liujiaxiang,luweixue,liyukun01,shidaiting01,wangshuaiqiang,chengzhicong01}@baidu.com}
\email{yindawei@acm.org}
\thanks{$^\#$ Co-first authors.}
\thanks{$^*$ Dawei Yin is the corresponding author.}

\begin{document}
\fancyhead{}
\pagenumbering{gobble} 

\begin{abstract}
Retrieval is a crucial stage in web search that identifies a small set of query-relevant candidates from a billion-scale corpus. Discovering more semantically-related candidates in the retrieval stage is very promising to expose more high-quality results to the end users. However, it still remains non-trivial challenges of building and deploying effective retrieval models for semantic matching in real search engine. In this paper, we describe the retrieval system that we developed and deployed in Baidu Search. The system exploits the recent state-of-the-art Chinese pretrained language model, namely Enhanced Representation through kNowledge IntEgration (ERNIE), which facilitates the system with expressive semantic matching. In particular, we developed an ERNIE-based retrieval model, which is equipped with 1) expressive Transformer-based semantic encoders, and 2) a comprehensive multi-stage training paradigm. More importantly, we present a practical system workflow for deploying the model in web-scale retrieval. Eventually, the system is fully deployed into production, where rigorous offline and online experiments were conducted. The results show that the system can perform high-quality candidate retrieval, especially for those tail queries with uncommon demands. Overall, the new retrieval system facilitated by pretrained language model (i.e., ERNIE) can largely improve the usability and applicability of our search engine. 
\end{abstract}

%
%



\maketitle

\section{Introduction}
Search engines (e.g., Google, Baidu, Bing) are critical tools for people to find useful information from massive web documents.
A modern search engine usually employ a multi-stage pipeline that gradually narrows down the number of relevant documents (i.e., web pages), where \emph{retrieval} is usually known as the very first stage. It aims at identifying a few hundreds or thousands of relevant candidates from the entire billion-scale corpus,
which has significant impact to the overall capability of a search engine.

Nevertheless, the unprecedented scale and diversity of web documents impose many challenges to the retrieval system. First (\textbf{C1}), \emph{semantic matching}~\cite{li2014semantic}
is one of the most critical concern, while conventional methods based on text matching (e.g., BM25~\cite{schutze2008introduction}) may easily fail at modeling relevant information with different phrasing.
Worse still, an increasing number of queries are in the style of natural language, making the semantic modeling even more challenging.
Second (\textbf{C2}), both search queries and web contents are highly heterogeneous, following long-tail distributions. For example, a large amount of low-frequency queries (i.e., tail queries) have never been seen before by the search engine. As such, the semantics of tail queries and documents are difficult to be accurately inferred.
Third (\textbf{C3}), to create significant impact to real-world applications, it also calls for practical solutions on deploying the retrieval model to serve web-scale data.

Extensive efforts from both academia and industry have been dedicated to tackle these challenges.
To conduct semantic retrieval, a wealth of studies~\cite{gao2011clickthrough,yih2011learning,salakhutdinov2009semantic,huang2013learning} have explored various bi-encoder models (i.e., Siamese networks or two-tower models), where different representation learning techniques has been employed as semantic encoders. Given a query and a document, the semantic encoder takes the query text and the document text (e.g., title) as inputs, and respectively produces two embeddings for the relevance computation.
Recently, BERT~\cite{devlin2018bert} has made significant progress on natural language understanding, and thus has also been applied as a more powerful semantic encoder~\cite{humeau2019poly,chang2020pre,lu2020neural}. The BERT-based bi-encoder and its variants have achieved the state-of-the-art performance on retrieval tasks~\cite{cai2021semantic,yates2021pretrained}, which can be mainly attributed to the expressive deep attention-based structure (i.e., Transformer) and the pretraining and fine-tuning paradigm. Although considerable research progress has been made, there is still a lack of investigation on how to develop and deploy such retrieval models in the online environment of a search engine.

\vspace{2mm}
\noindent\textbf{Present work.} In this paper, we introduce a novel retrieval system that is fully deployed in Baidu Search\footnote{www.baidu.com}, the dominant search engine in China. In particular, the retrieval system is equipped with the recent state-of-the-art Chinese pretrained language model (PLM), namely \emph{Enhanced Representation through kNowledge IntEgration (ERNIE)}~\cite{sun2019ernie}, which enables effective semantic matching in the system. More specifically, we design the retrieval system with several key insights to tackle the aforementioned challenges:

\begin{itemize}[leftmargin=*]
\item To tackle the first challenge (\textbf{C1}), we leverage an ERNIE-based (i.e., Transformer-based) bi-encoder to perform expressive semantic matching. In particular, the Transformer encoders directly take the raw texts of queries and web documents as inputs and encode their semantics in latent embeddings. The deep structure of Transformer encoders allow the complicated semantics to be more comprehensively modeled. The dense attention over the raw texts can also keep the semantics of fine-grained context, such as using different prepositions (e.g., ``for'' vs. ``to''),
which is more friendly to conversational queries.
Moreover, we integrate a poly-interaction scheme~\cite{humeau2019poly} and effective training data mining strategies, which further improves the effectiveness of the retrieval model.

\item To tackle the second challenge (\textbf{C2}), we further propose a multi-stage training paradigm for optimizing the retrieval model. In particular, the training stages are designed with different data sources and objectives, which allows rich knowledge to be absorbed by the model. Compared with training-from-scratch, our proposed paradigm can further boost the generalization ability of the model, which is especially beneficial for tail queries.

\item To tackle the third challenge (\textbf{C3}), we develop a system architecture that serves the proposed model for large-scale web retrieval. In particular, we incorporate the semantic matching with conventional text matching to collaboratively generate relevant candidates. Moreover, we further introduce a lightweight post-retrieval filtering module that provides an unified filtering for the retrieval, where more statistical features (e.g., clickthrough rate, dwell time) can be introduced to consider the overall quality. Overall, the system architecture allows the proposed model to work together with conventional text-matching workflow and offers flexibility for including other features.
\end{itemize}

To the best of our knowledge, this is one of the largest applications of pre-trained language models for web-scale retrieval. 
We anticipate this paper to provide our practical experiences and new insights for incorporating PLMs in web retrieval. 
The main contributions can be summarized as follows:
\begin{itemize}[leftmargin=*]
    \item \textbf{The retrieval model.} 
    We leverage Transformers as the semantic encoders for web retrieval, and further exploit a poly-interaction scheme and several strategies for mining training data.
    The model can capture complicated semantic information underlying query and web documents, and is effective for semantic matching.
    \item \textbf{Training paradigm.} We introduce a novel multi-stage training paradigm that facilitates the retrieval model to learn rich information. This would significantly improve the generalization ability of the model, especially for the tail queries.
    \item \textbf{System design}. We design an effective and efficient system workflow to server the model, allowing it to seamlessly work with the conventional workflow to boost the overall performance. We also introduce a post-retrieval filtering module, which integrates statistical features and semantic relatedness that measure the overall quality of the candidates in an unified manner.
    \item \textbf{Evaluation}. We conduct extensive offline and online experiments to validate the effectiveness of the retrieval system. The results show that the deployment of the system can significantly improve the usability and applicability of the search engine.
\end{itemize}

\section{Related Work}
\subsection{Semantic Retrieval in Web Search}
Semantic retrieval is essential for a modern retrieval system. Typical structures of semantic retrieval models can be viewed as bi-encoders or Siamese networks~\cite{dong2018triplet},
which comprises two encoders that conduct semantic modeling.
Most of the existing studies mainly focus on designing the encoders with different representation learning techniques ~\cite{gao2011clickthrough,yih2011learning,salakhutdinov2009semantic,huang2013learning,huang2020embedding,liu2020decoupled}.
A representative work, namely Deep Semantic Similarity Model (DSSM)~\cite{huang2013learning}, is one of the earliest DNN methods for semantic modeling with clickthrough data. The deep fully-connected architectures of the DSSM encoders can extract expressive semantic information and achieve superior performance on web search. 
Thereafter, other DNN-based retrieval methods have thrived over the past years~\cite{mitra2018introduction}, including those based on Convolutional Neural Networks (CNNs)~\cite{shen2014learning,gao2019modeling,shen2014latent,severyn2015learning} and Recurrent Neural Networks (RNNs)~\cite{palangi2015deep,palangi2014semantic}. 

The common bi-encoder structure of those models allows a large number of document embeddings to be pre-computed and cached offline, which is very efficient for online retrieval. Therefore, such model structure has also been developed in real-world products~\cite{huang2020embedding,zhang2020towards,haldar2020improving}. For instance, Huang \etal~\cite{huang2020embedding} employ DNN-based bi-encoder in Facebook Search system, and introduce various practical experiences in the end-to-end optimization of the system. Zhang \etal~\cite{zhang2020towards} introduce embedding learning for the semantic retrieval in their E-commerce search service. Besides, Yin \etal~\cite{yin2016ranking} also adopts deep semantic matching in the core ranking module of Yahoo Search. Different from previous studies, 
we explore pretrained language model with multi-stage training to better perceive the semantics behind queries and documents.

Despite the above-mentioned bi-encoder models, \emph{interaction-based methods} are also widely used in many information retrieval systems~\cite{gu2020deep,zou2019reinforcement,zou2020neural,zou2020pseudo,guo2016deep,zhao2020autoemb,zhao2020memory,yates2021pretrained}.
As such, another line of research for semantic matching is to model query-document \emph{interaction} with DNNs~\cite{yates2021pretrained,mitra2017learning,lu2013deep,wan2016deep,zou20201pretrained}. However, they cannot cache the document embeddings offline, and thus are inefficient for retrieval. They are preferred for ranking stage, which will not be further discussed in this paper. In our search engine, interaction-based methods are exploited to build the PLM-based ranking system~\cite{zou20201pretrained}.

\subsection{Pretrained Language Models}
Pretrained Language Models (PLMs), such as ELMo~\cite{peters2018deep}, BERT~\cite{devlin2018bert} and ERNIE~\cite{sun2019ernie}, have achieved monumental success in natural language understanding. Notably, the recent state-of-the-art PLMs~\cite{devlin2018bert,liu2019roberta,yang2019xlnet} are usually based on Transformers~\cite{vaswani2017attention}, which exploit a deep structure with stacked multi-head attention and fully-connected layers. More importantly, they adopted unsupervised pretraining with large corpus, and thus can incorporate more useful knowledge in the models.
As BERT and its successors (e.g., XLNet~\cite{yang2019xlnet}, RoBERTa~\cite{liu2019roberta}) exhibit superior capacity in understanding textual data, a handful of studies start to leverage them for semantic retrieval~\cite{chang2020pre,humeau2019poly,lu2020twinbert}. For instance, Chang \etal~\cite{chang2020pre} propose an early attempt that introduces BERT-based bi-encoders for large-scale retrieval. It also studies the effects of several new pretraining tasks. Humeau \etal~\cite{humeau2019poly} advance such bi-encoder with attentive interaction for the query and document embeddings. Lu~\etal~\cite{lu2020neural} explore the negative sampling strategies for BERT-based bi-encoders at both pretraining and finetuning stages. However, despite the initial success achieved by these work, there is still a lack of investigation on developing and deploying such model for web-scale retrieval, especially in real-world productions. In this paper, we propose to leverage the state-of-the-art Chinese pretrained language model, namely Enhanced Representation through kNowledge IntEgration (ERNIE)~\cite{sun2019ernie}, for building the retrieval system of our search engine.

\section{Retrieval Model}
In this section, we
first present the task definition, and then introduce the details of our proposed retrieval model.

\subsection{Task Definition}
Given a query $q$,
the web retrieval task aims to return a set of relevant documents from a large corpus $D$, where the size of the corpus can be tens of billions or more. We denote by $D_q^+$ the set of documents relevant to the query. 
Different from conventional retrieval with term matching,
\emph{semantic retrieval} computes the relevance score $f(\cdot)$ in a learned embedding space \cite{lee2019latent,luan2020sparse,karpukhin2020dense,xiong2020approximate,chang2020pre}, using similarity metrics (e.g., dot-product, cosine similarity) as:
\begin{equation}\label{eq:bi-encoder}
    f(q,d) = \mathrm{sim}(\phi(q; \theta), \psi(d; \beta)),
\end{equation}
where $\phi$ (or $\psi$) is the representation model (i.e., encoder) parameterized by $\theta$ (or $\beta$) that encodes the query (or document) to dense embeddings. 

\subsection{Model Architecture}\label{sec:model_arch}
\begin{figure}[!t]
    \centering
    \includegraphics[width=0.45\textwidth]{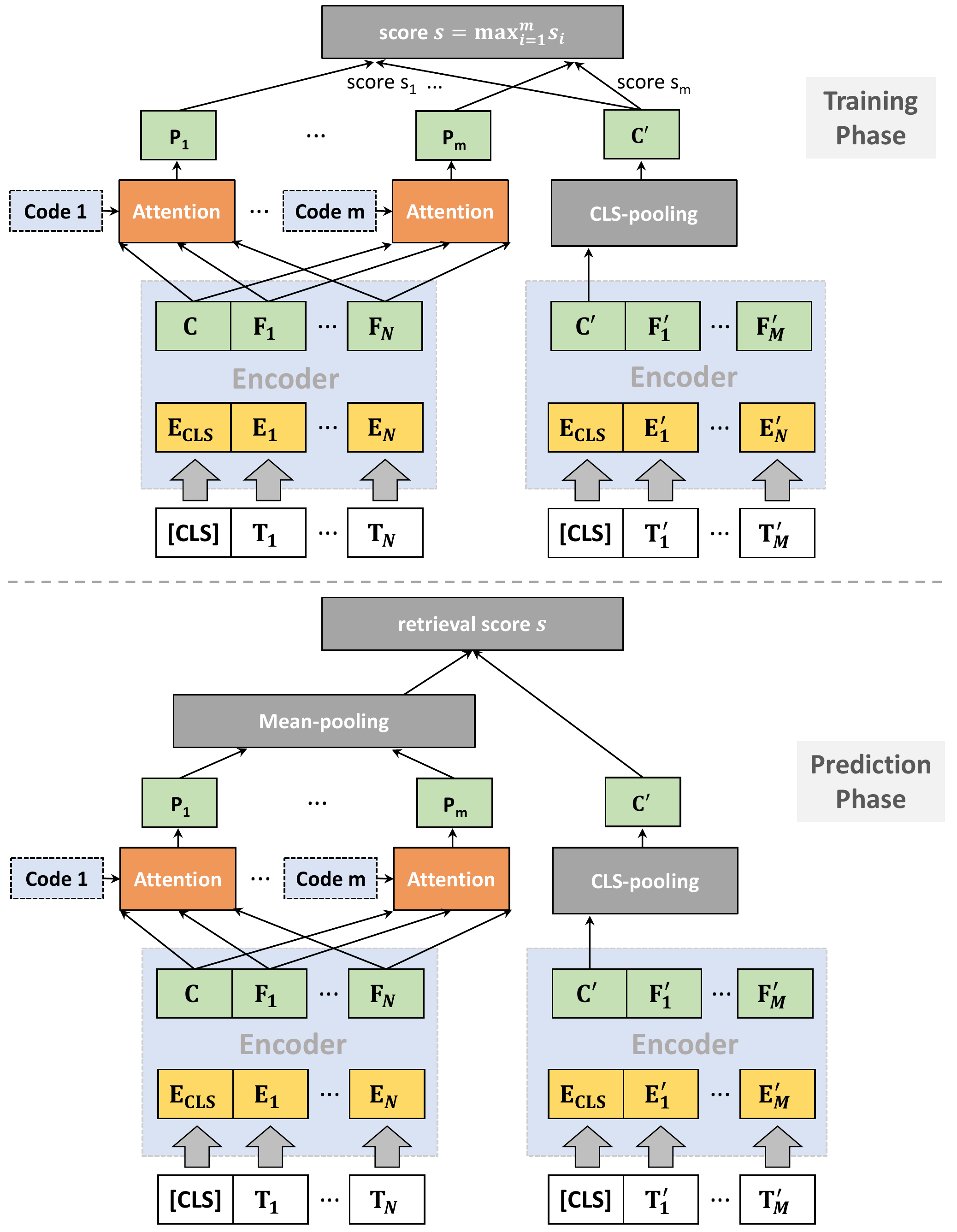}
    \caption{Model architectures and relevance computation.}
    \label{fig:poly-encoder}
\end{figure}

In this work, we are interested in parameterizing the encoders $\phi$ and $\psi$ as deep Transformer models \cite{vaswani2017attention} due to its expressive power in modeling natural language.

\noindent\textbf{The bi-encoder backbone}.
By implementing $\phi$ and $\psi$ as deep Transformers, we refer the paradigm described in Eq. (\ref{eq:bi-encoder}) as bi-encoder \cite{humeau2019poly}. More specifically, let $\mathcal{T}$ be a multi-layer transformer encoder, which is a stack of $n$ transformer blocks. Each block consists of a multi-head self-attention (MHA) sublayer followed by a feed-forward (FFN) sublayer, where MHA allows the model to jointly attend to different subspaces and FFN aggregates the attention results of different heads.  
The encoder takes $\left\{ [\mathrm{CLS}], T_1, \cdots, T_N  \right\}$ (or $\left\{ [\mathrm{CLS}], T^{\prime}_1, \cdots, T^{\prime}_M  \right\}$), i.e., the tokenzied sequence of the raw query (or document) text, as the input and outputs an encoded sequence $C, F_1, \cdots, F_N$ (or $C', F'_1, \cdots, F'_M$) as described in Figure \ref{fig:poly-encoder}. Note that [CLS] is a pseudo token that aggregates information in the encoder for the subsequent matching. 
For each transformer layer, the parameters are shared between query and document encoders (i.e., $\phi$ and $\psi$),
which has multiple potential benefits, such as reducing the number of model parameters so as to control model complexity \cite{firat2016zero}, introducing prior knowledge to regularize models \cite{xia2018model}, and saving the storage space or memory size \cite{johnson2017google}. The conventional bi-encoder computes the relevance score with a simple dot-product between the output CLS embeddings, i.e., $f(q, d)= C\cdot C^{\prime}$.

\noindent\textbf{Poly attention}.
Different from vanilla bi-encoder, we further develop a poly attention scheme that enables more flexibility for modeling query semantics. This idea is originated from poly-encoder~\cite{humeau2019poly} and is slightly customized in our retrieval model, which works differently during training and prediction phases.

In the training phase, as shown in the upper half part of Figure \ref{fig:poly-encoder}, we introduce a set of $m$ context codes, i.e., $c_0, \cdots, c_{m-1}$, where each $c_i$ extracts a global representation $P_i$ by attending over all the outputs of the query encoder (i.e., $C, F_1, \cdots, F_N$) as
\begin{equation}
    P_i =  w_0^{(i)} \cdot C + \sum_{j=1}^{N} w_j^{(i)} \cdot F_j,
\end{equation}
where $\left( w_0^{(i)}, \cdots, w_{N}^{(i)} \right) = \mathrm{Softmax}\left( c_i \cdot C, \cdots, c_i \cdot F_{N} \right)$. These global context features $P_1, \cdots, P_m$ can be interpreted as different aggregations of the semantics from all the query terms, which reflects fine-grained query demands.
Subsequently, each of the $m$ global context features is compared with the document representation $C'$ using dot-product, followed by a max-pooling to finalize the relevance score, i.e., 
\begin{equation} \label{eq:poly_score_train}
    f(q, d) = \mathrm{max}_{i=1}^{m} P_i \cdot C'.
\end{equation}

In the retrieval/prediction phase, we employ a slightly different strategy due to a practical concern. In real application with massive web documents, document representations must be pre-computed to build indexes, which enable efficient retrieval with fast nearest neighbor search. In such case, relevance computation based on Eq. (\ref{eq:poly_score_train}) is clearly infeasible, as such metric with max-pooling can hardly be supported by existing index schemes. Motivated by this, we propose to construct an unified surrogate embedding with mean pooling over all $P_i$, i.e., $\frac{1}{m}\sum_{i=1}^{m} P_i$. As such, the final relevance score is defined as 
\begin{equation}  \label{eq:poly_score_pred}
f(q,d) = P_{avg} \cdot C' = \left( \frac{1}{m}\sum_{i=1}^{m} P_i \right) \cdot C'.  
\end{equation}
With such definition, index-based nearest neighbor search can be leveraged, which largely improves the applicability of the model.

\noindent\textbf{Remark.} As already mentioned, the relevance score is calculated inconsistently between training and predicting phases. In practice, we find out that such inconsistency would not undermine the model performance for semantic retrieval. Therefore, through bagging distinct features learnt by the context codes, we can achieve a better and more heterogeneous representation of the query for a more powerful semantic matching with that of the document.

\subsection{Optimization}\label{sec:loss_function}
We formulate the learning procedure of the retrieval model using maximum likelihood estimation: Given a query $q$, all of its relevant documents $D_q^+$ and irrelevant ones $D_q^-$, the optimal parameters $\theta^\ast, \beta^\ast$ can be defined as:
\begin{equation}\label{eq:optimization}
    \theta^\ast, \beta^\ast = 
    \arg \max_{\theta, \beta} 
    \sum_{q} \sum_{d^+ \in D_q^+} 
    p(d^+|q, d^+, D_q^-),
\end{equation}
where $p(d^+|q, d^+, D_q^-)$ is the probability that $f$ can separate a relevant document $d^+$ from all irrelevant ones $D_q^-$. Note that we consider the context codes $c_i$ are the parameters of the query encoder, which are jointly trained. 

We implement $p(d^+|q, d^+, D_q^-)$ as a contrastive probability, i.e.,
\begin{equation}
    p(d^+|q, d^+, D_q^-) = \frac{ \exp \left( f(q, d^+) / \tau \right) }
        { \displaystyle \exp \left( f(q, d^+) / \tau \right) + \sum_{d^- \in D_q^-} \exp \left( f(q, d^-) / \tau \right)  },
\end{equation}
where $\tau$, which is normally set to 1, is the temperature of the softmax operation~\cite{hinton2015distilling}. Using a higher value for $\tau$ produces a softer probability distribution over classes (i.e., relevant or irrelevant). 

Furthermore, as it is inapplicable to know and use all the real relevant and irrelevant documents in the training,
we reformulate the optimization problem with sampled relevant and irrelevant documents (respectively denoted as $\hat{D}_q^+$ and $\hat{D}_q^-$) as
\begin{equation}
\label{eq:sampled_optimization}
\theta^\ast, \beta^\ast = 
\arg \max_{\theta, \beta} 
\sum_{q} \sum_{d^+ \in \hat{D}_q^+} 
p(d^+|q, d^+, \hat{D}_q^-).
\end{equation}

\begin{figure}[!t]
    \centering
    \includegraphics[width=0.45\textwidth]{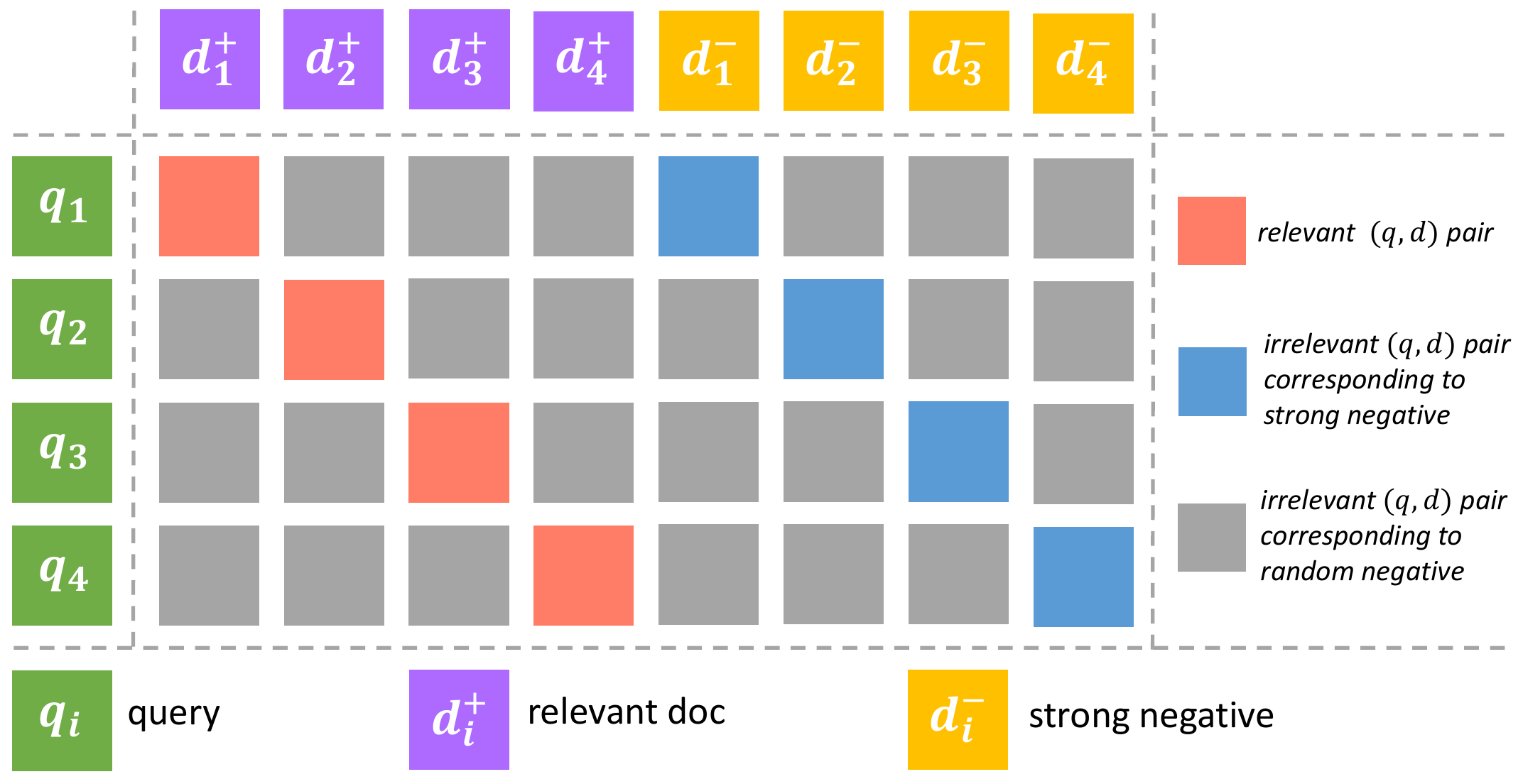}
    \caption{Illustration of negative sampling during training.}
    \label{fig:negatives}
    \vspace{-2mm}
\end{figure}

\subsection{Training Data Mining} \label{sec:data_mining}
To solve the optimization problem as Eq. (\ref{eq:sampled_optimization}), the key of model training lies on the construction of $\hat D_q^+$ and $\hat D_q^-$, such that a better model can be learned for semantic retrieval. In this subsection, we elaborate how we mine the positive and negative data for training.

\noindent\textbf{Positives and negatives in different data sources}.
For mining positive and negative examples, we first consider two kinds of data sources that are commonly used in practice:
\begin{itemize}[leftmargin=6mm]
    \item \emph{Search log data}, where the queries and documents are logged with click signal, i.e., whether the document is clicked by the user. For each query, we use those clicked results as positives and those exposed but non-click results as negatives, as clicks can roughly indicate the satisfaction of user's intent.
    \item \emph{Manually labeled data} is usually collected with fine-grained grades (i.e., 0 to 4) assigned by human evaluation.
    For each query, the positive and negative examples are defined in a pairwise manner. For a given document that is considered as positive, other lower-grade documents under the same query can be considered as negatives.
\end{itemize}

\noindent\textbf{In-batch negative mining}. 
The goal of online retrieval is to distinguish a tiny fraction of relevant documents from massive irrelevant documents (i.e., $|D_q^+| \ll |D_q^-| \approx |D|$). Instead of logging (or labeling) many irrelevant documents, we leverage an in-batch negative mining scheme~\cite{karpukhin2020dense}, which is a more efficient way to construct those completely irrelevant documents. To avoid confusion, we refer to the non-click (or lower-grade) samples as \emph{strong negatives}, and the in-batch negatives as \emph{random negatives}. 

Particularly during the training of our model, we consider random negatives as those documents (i.e., positives and strong negatives) of other queries in the same mini-batch. As data are shuffled before fed to the model, the random negatives are generally quite distinct from the query, which helps mimic the online retrieval scenario, i.e., identifying positives from massive random negatives.
In addition, to collect sufficient random negatives from each mini-batch, we exploit a simple yet practical solution, i.e., increasing the batch size. We apply mix-precision method that allows larger batch size during training with a fixed memory consumption. The batch size is set to make maximum use of GPU memory.

Figure \ref{fig:negatives} shows an example of the overall negative sampling strategies in a more intuitive way. In the example, the mini-batch of size $4$ consists of four triplets $\{ (q_i, d^+_i, d^-_i), i=1,2,3,4 \}$. For the query $q_1$, the positive and strong negative documents are respectively denoted as $d^+_1$ and $d^-_1$, and $d^+_2, d^+_3, d^+_4, d^-_2, d^-_3, d^-_4$ are 
random negatives. For each row (i.e., each query) in Figure~\ref{fig:negatives}, there is one positive and a set of negatives, which can be used as $d_+$ and $D_q^-$ in Eq. (\ref{eq:sampled_optimization}). Therefore, the training data mined with these strategies can be used to optimize the retrieval model based on Eq. (\ref{eq:sampled_optimization}).

\begin{figure*}[!t]
    \center
    \includegraphics[width=\textwidth]{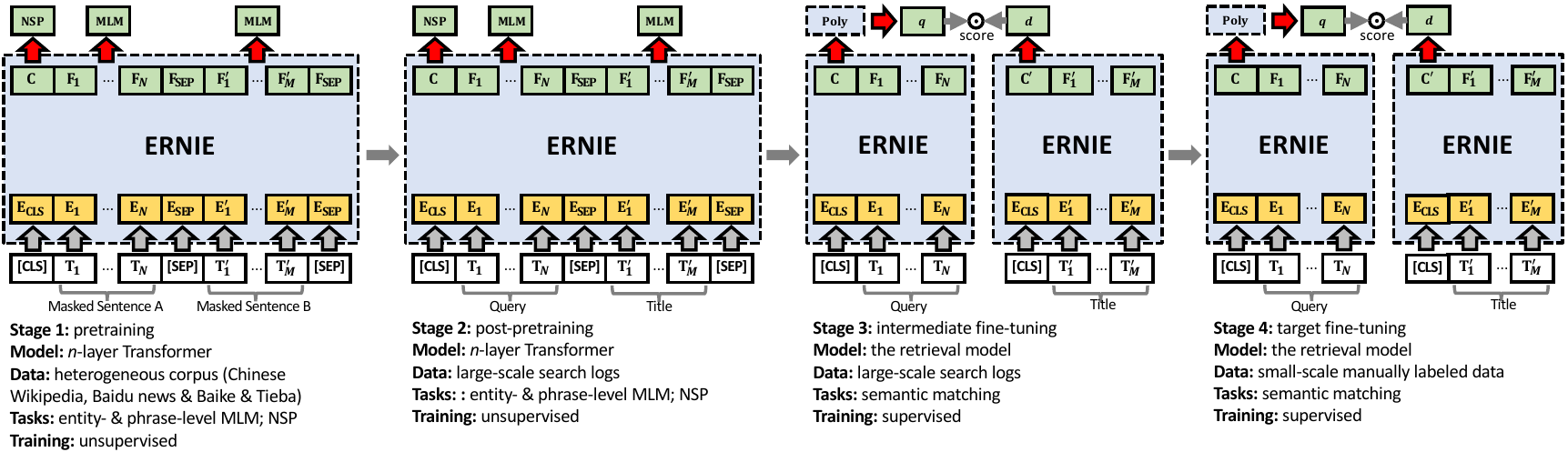}
    \caption{Training Paradigm of ERNIE-based Retrieval Model.}
    \label{fig:pipeline}
    \vspace{-2mm}
\end{figure*}

\section{Training Paradigm} \label{sec:training}

The training paradigm of pretraining and finetuning has been widely employed for model optimization in many natural language processing problems. Representations learned by such paradigm usually show competitive performance across many tasks. However, such superiority has not yet been fully exploited for web retrieval.
In this section, we propose a novel multi-stage training paradigm. Particularly, as shown in Figure \ref{fig:pipeline}, we divide the entire training process into four stages: (1) \textbf{pretraining}, (2) \textbf{post-pretraining}, (3) \textbf{intermediate fine-tuning}, and (4) \textbf{target fine-tuning}. Each stage is carefully designed with different training data and objectives, which boost the generalization ability of the model. The overall training pipeline is shown in Figure \ref{fig:pipeline}.

\subsection{Stage 1: Pretraining}
In this stage, we follow the same pretraining process in \cite{sun2019ernie} to train an ERNIE encoder (i.e., an $n$-layer transformer). It adopts heterogeneous corpus, i.e., Chinese Wikipedia, Baidu Baike (containing encyclopedia articles written in formal languages), Baidu news and Baidu Tieba (an open discussion forum like Reddits), to learn language representation enhanced by entity- and phrase-level masking strategies. The heterogeneous corpus contributes a large proportion of web documents in our search engine, hence the pretrained ERNIE model can effectively transfer knowledge to web retrieval. The pretraining is unsupervised, where the tasks include masked word prediction (i.e., MLM - Masked Language Modeling) and the next sentence prediction (NSP).

\subsection{Stage 2: Post-pretraining}
Search engine processes billions of various queries with diverse goals or intentions every day, such as medical advice, travel guides, latest news, etc. Pretraining only on document corpus might be limited to tackle all kinds of requests from users. In this stage, we transfer previous pretrained model and continue pretaining on both query and document
corpus of Baidu Search with the same tasks (i.e., MLM and NSP) as Stage 1. 

We name Stage 2 as post-pretaining, which keeps the model structure (i.e., a $n$-layer Transformer) and the training tasks while changes the training data.
Specifically, we collect one-month (i.e., tens of billions of) user search logs
for post-pretraining. 
The tokenized raw texts of the query and the title of the document are concatenated as the input during this stage. Then we apply the same masking strategies as ERNIE for MLM and take clicked documents as the next sentence of the query for NSP.

\subsection{Stage 3: Intermediate Fine-tuning}
From this stage, we start to fine-tune the retrieval model. We first fine-tune the retrieval model on an intermediate task, before fine-tuning again on the target task of interest \cite{pruksachatkun2020intermediate}. 
In particular, we leverage the post-pretrained ERNIE produced by Stage 2, as the encoders of our retrieval model, which is subsequently optimized using the same training data (i.e., search logs) as Stage 2. The training objective is as presented in Eq. (\ref{eq:sampled_optimization}), and we construct positives and negatives for training as introduced in Section \ref{sec:data_mining}.
Using an intermediate objective that more closely resembles our target task leads to better and faster fine-tuning performance.

\subsection{Stage 4: Target Fine-tuning}
Finally, we fine-tune our retrieval model produced by Stage 3 on a relatively small manually-labeled data, which we consider is the most accurate and unbiased data for learning the retrieval model. For the data collection, a set of queries are sampled from query logs, and a certain number of query-document pairs are labeled according to their relevance judged by human editors. A 0-4 grade is assigned to each query-document pair based on the degree of relevance. The training objective is as presented in Eq. (\ref{eq:sampled_optimization}), and we construct positives and negatives for training as introduced in Section \ref{sec:data_mining}.

\begin{figure*}[!t]
    \centering
    \includegraphics[width=0.9\textwidth]{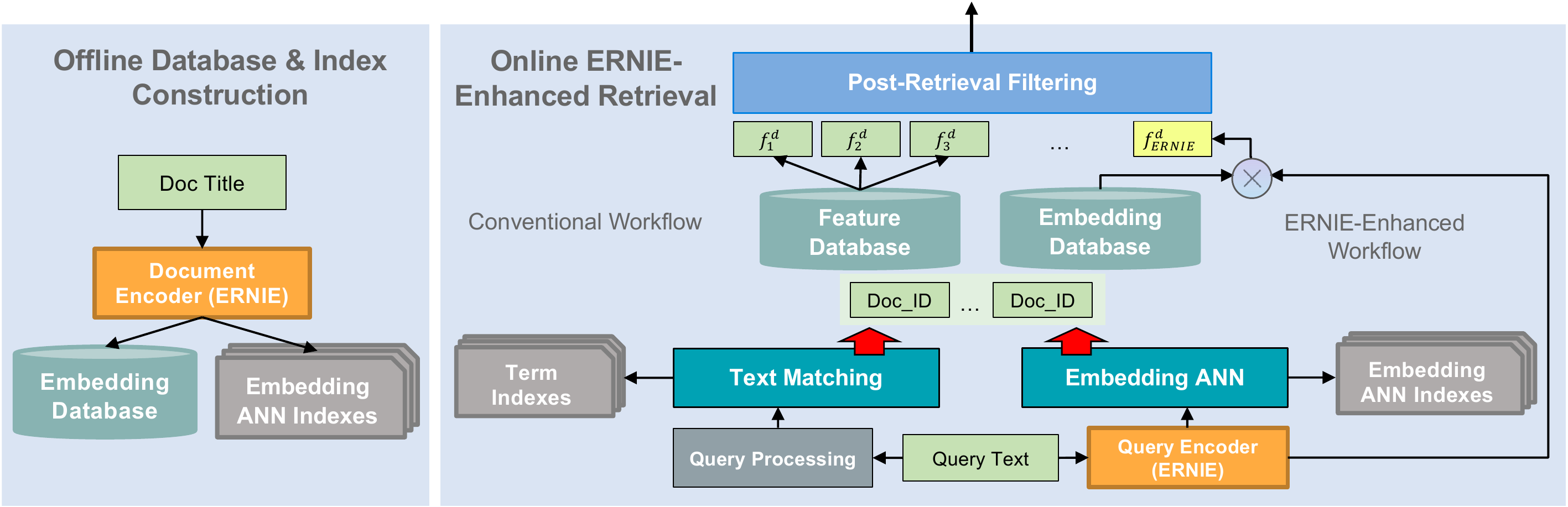}
    \caption{The overall workflow of the ERNIE-enhanced retrieval system.}
    \label{fig:sys_arch}
\end{figure*}

\section{Deployment}
In this section, we first introduce the embedding compression and quantization, which saves the online cost for retrieval. Then, we present the general picture of how the retrieval model works in the retrieval system. For the detailed implementation of our deployed model, please refer to the Appendix.

\subsection{Embedding Compression and Quantization }
We apply compression and quantization for the output embeddings of the retrieval model. On the document side, it can significantly reduce the memory cost for caching the embeddings; On the query side, it saves the transmission cost for query embedding, and thus improves the system throughput. 

\noindent\textbf{Compression}.
We jointly train a compression layer with the ERNIE encoders. The compression layer is implemented as a fully-connected layer that takes the output embeddings of the encoders (i.e., $P_i$ and $C'$) as inputs, and produces lower-dimensional embeddings. This would largely save the memory cost, with very slight decrease w.r.t. accuracy. In practice, we reduce the size of output embeddings from 768 to 256, which improves the overall efficiency of transmission and storage by 3 times.

\noindent\textbf{Quantization}.
Quantization is a very promising technique for boosting the efficiency and scalability of neural networks~\cite{hubara2017quantized}. For deploying our model to efficiently serving massive search queries, we apply a commonly-used int8 quantization to the output embeddings, where the overall efficiency can further be boosted by 4 times. More details of the embedding quantization can be referred to the Appendix.

\subsection{System Workflow}
An important fact for building an effective retrieval system is that, neither explicit text matching nor embedding-based semantic matching can handle all the various kinds of queries. Instead, it is more promising to integrate the two types of retrieval methods together, to provide a better overall performance.
Motivated by this, we developed a novel system workflow, which is depicted in Figure \ref{fig:sys_arch}. Compared with conventional text matching, the system is upgraded with two modules:
\begin{itemize}[leftmargin=6mm]
    \item \emph{ERNIE-based retrieval}. The system enables the ERNIE-based retrieval model to work in parallel with the conventional retrieval workflow (i.e., text matching), which offers high maintainability of the system.
    \item \emph{ERNIE-enhanced post-retrieval filtering}. We further introduce an unified post-retrieval filtering module. It takes both retrieved documents from text matching and ERNIE-based semantic matching, and conduct effective post-retrieval filtering for further discovering high-quality candidates.
\end{itemize}
We elaborate the system workflow with the two modules, which including both offline and online stages.

\noindent\textbf{Offline database and index construction}.
As shown in the left part of Figure \ref{fig:sys_arch}, during the offline stage, we use the trained ERNIE encoder to compute the embeddings (with the compression and quantization) for all documents in the corpus, based on which we build 1) Approximate Nearest Neighbor (ANN) indexes, and 2) an embedding database for the above-mentioned two modules, respectively. In particular, the ANN indexes are deployed for the ERNIE-based retrieval stage, to enable efficient embedding-based candidate generation via fast nearest neighbor search. The embedding database is used for the post-retrieval filtering stage. It is a key-value database for efficient document embedding lookup with given document ids. Next, we introduce the online workflow and show how the database and indexes are used online.

\noindent\textbf{Online integrated retrieval}.
For the online retrieval, we integrate conventional retrieval and the novel ERNIE-based semantic retrieval. The conventional retrieval first processes the query text with several operations (e.g., word segmentation, stop-word filtering), and then conducts keyword/term matching with term indexes (i.e., inverted indexes) to retrieve a set of documents. In parallel, semantic retrieval first fetch the query embedding. After online embedding compression and quantization, the query embedding is applied for semantic retrieval with the ANN indexes. The results produced by different retrieval approaches are merged to form a candidate pool.

\noindent\textbf{Online post-retrieval filtering}.
After the integrated retrieval, we further introduce an online post-retrieval filtering stage, which further narrows down the scale of the candidates. To achieve this, the filtering module contains a lightweight ranking model (e.g., GBRank or RankSVM) that conducts an unified comparison for the retrieval candidates. Particularly for each candidate document, we fetch the statistical features (denoted as $f^d_1, f^d_2, f^d_3, ...$ in Figure \ref{fig:sys_arch}), such as click-through rate, from the feature database. Note that the semantic matching score is also applied in this stage as an important feature (denoted as $f^d_{ERNIE}$), where the scores computed in semantic retrieval can be reused. For those candidates retrieved by text matching, where the semantic matching scores are missing, we fetch their embeddings from the embedding database, and then compute the scores with the query embedding. As the candidate pool formed by the online retrieval stage is small compared to the entire corpus, the online computation of the semantic matching scores does not significantly increase the time cost. By doing this, we unify the comparison of all candidate documents with several statistical features and a semantic feature, where the semantic modeling capacity of our retrieval model can also benefit the documents dug by conventional retrieval paradigm.

\section{Offline Evaluation}
We conduct an offline evaluation for the proposed retrieval model. 
The implementation details of the model and more ablation study can be found in the Appendix.

\subsection{Datasets} \label{sec:datasets}
We evaluate the retrieval model on the datasets that are collected from the real production environment. Note all the data does not contain any user-related information for the privacy concern.

\noindent\textbf{Manually-labeled dataset (Manual)}.
This dataset contains 6,423 queries and 147,341 query-document pairs. Each query has around 22 retrieved results on average. For each query, we collect documents from each stage of the search pipeline to ensure the diversity of the dataset. The dataset is labeled on our crowdsourcing platform, where a group of experts are required to assign an integer score varies from 0 to 4 to each query-document pair. The score represents whether the content of the document w.r.t. the query is off-topic (0), slightly relevant (1), relevant (2), useful (3) or vital (4).
    
\noindent\textbf{Automatic-annotated datasets (Auto)}. We collect another two search log datasets, namely Auto-Rand and Auto-Tail, which contains random queries and tail queries, respectively. Here, the tail queries are identified as those who have no more than 10 searches per week. For each query, we use our search engine to collect the top-10 results in the final ranking stage as the positive examples (i.e., the relevant documents), and consider the top-10 results of other queries as negative examples. After filtering noise and meaningless queries, we finalize the two datasets, where Auto-Rand contains 18,247 queries and 112,091 query-document pairs, and Auto-Tail contains 78,910 queries and 750,749 query-document pairs.

\subsection{Evaluation Metrics}
\noindent\textbf{Positive-Negative Ratio}.
We report the positive-negative ratio (PNR) on Manual dataset. For a given query $q$ and the associated documents $D_q$, the PNR can be formally defined as 
\begin{equation}
    PNR= \frac{\sum_{d_i, d_j \in D_q} \mathbbm{1}(y_i > y_j)\cdot \mathbbm{1}(f(q, d_i) > f(q, d_j))}{\sum_{d_{i^\prime}, d_{j^\prime} \in D_q} \mathbbm{1}(y_{i^\prime} > y_{j^\prime})\cdot \mathbbm{1}(f(q, d_{i^\prime}) < f(q, d_{j^\prime}))},
\end{equation}
where $\mathbbm{1}(\cdot)$ is an indicator function, i.e., $\mathbbm{1}(a > b)=1$ if $a > b$, and 0 otherwise.
PNR measures the consistency between the manual labels and the model scores. We report the averaged PNR values over all queries in the experiments.

\noindent\textbf{Recall@10}. We report Recall@10 on Auto-Rand and Auto-Tail datasets. The metric is defined as $Recall@10 = |D_q \cap \tilde D_q| / 10$,
where $\tilde D_q$ represents the top-10 retrieval w.r.t. $q$ using the retrieval model considering all the documents in a dataset as the corpus, and $D_q$ is the set of collected ground truth. We report the averaged Recall@10 values over all queries in the experiments.

\begin{table}[t]
\caption{Offline experimental results on different stages.}
\vspace{-2mm}
\scalebox{0.93}{ \label{exp:offline}
\begin{tabular}{l|c|cc}
\hline\hline
Dataset & Manual  & Auto-Rand & Auto-Tail \\ \hline
Metric & PNR  & Recall@10 & Recall@10 \\ \hline
Pretrain                   & 1.34  & 18.92\%          & 12.10\%                \\
Post-pretrain                    & 1.43  & 35.99\%          & 19.10\%                \\
Intermediate Fine-tune                   & 2.00 & 83.14\%          & 60.98\%                \\
Target Fine-tune                  & \textbf{2.48} & \textbf{87.90\%}          & \textbf{71.13\%}          \\ \hline
Online baseline                         & 1.77 & 85.22\%          & 65.79\%          \\ \hline \hline
\end{tabular}}
\end{table}

\subsection{Offline Experimental Results}
In the offline evaluation, we report the experimental results of the proposed model during different training stages. Here, the performance of pretraining and post-pretraining are reported when using the $n$-layer Transformer in a bi-encoder model. Besides, we also include a baseline method, i.e., the model that is used in the system before deploying the ERNIE-base retrieval model. 

Table \ref{exp:offline} shows the results, where some key findings are observed:
\begin{itemize}[leftmargin=6mm]
    \item The performance on different stages indicate that our proposed training paradigm is able to gradually improve the performance of the retrieval model. In particular, we can see that the model in the pretrain stage does not contain any domain knowledge w.r.t. the retrieval tasks, and thus performs poorly, where the Recall@10 is below 20\% on both Auto datasets. However, after applying the multi-stage training paradigm, the Recall@10 of the model finally reaches 87.90\% and 71.13\% on Auto-Rand and Auto-Tail, respectively.
    \item By applying the multi-stage training, the retrieval model can outperforms our online baseline w.r.t. Recall@10 on both Auto datasets. Moreover, we also observe that the relative improvement of our proposed model over the online base is large on Auto-Tail ($\Delta=8.1\%$) than Auto-Rand ($\Delta = 3.1\%$), in terms of Recall@10. This shows that our proposed method can better discover relevant results for tail queries, which is very helpful in improving the user experience for the search engine.
    \item In addition, we can see that the proposed model can beat the online baseline by a large margin w.r.t. PNR, where the value is improved from 1.77 to 2.48. This tells us that the proposed model not only can retrieval relevant documents, but also prefer high-quality results, i.e., the documents with higher manually-labeled grades.
\end{itemize}
Overall, our proposed model is able to gain superior performance on semantic retrieval through the multi-stage training, and beat the online baseline by a significant margin.

\begin{table*}[t]
\caption{Interleaved comparison.} \label{exp:interleaving}
\vspace{-2mm}
\scalebox{+0.93}{
\begin{tabular}{l|cc|cc|cc|cc}
\hline \hline
\multirow{3}{*}{} & \multicolumn{4}{c|}{ENINE-based retrieval}                         & \multicolumn{4}{c}{Post-retreival filtering}                         \\ \cline{2-9} 
                  & \multicolumn{2}{c|}{Query type}   & \multicolumn{2}{c|}{Query length} & \multicolumn{2}{c|}{Query type}   & \multicolumn{2}{c}{Query length} \\ \cline{2-9} 
                  & Rand    & Tail (freq $<3$) & Short ($\leq$10)  & Long ($10<$)  & Rand    & Tail (freq $<3$) & Short ($\leq$10)  & Long ($10<$)  \\ \hline
$\Delta_{AB}$     & +0.368\% & +0.992\%                 & +0.345\%           & +0.426\%       & +0.112\% & +0.274\%                 & +0.066\%           & +0.191\%       \\ \hline
$\Delta_{AB}$-tw  & +0.281\% & +0.783\%                 & +0.253\%           & +0.352\%       & +0.226\% & +0.454\%                 & +0.176\%           & +0.315\%       \\ \hline \hline
\multicolumn{9}{c}{*All the values are statistically significant ($t$-test with $p<0.05$).}
\end{tabular}}
\vspace{-2mm}
\end{table*}

\section{Online Evaluation}
To investigate the impact of our proposed system to the search engine, we deploy the new system and conduct online experiments to compare it with the old retrieval system. 

\subsection{Interleaved Comparison}
\emph{Interleaving}~\cite{chuklin2015comparative} is a commonly-used technique for evaluating industrial information retrieval systems (e.g., recommender systems, search engines). In interleaved comparison, the results of two systems are interleaved and exposed together to the end users, whose click behaviors would be credited to the system that provides the clicked results. The gain of the new system A over the base system B can be quantified with $\Delta_{AB}$, which is defined as 
\begin{equation}\label{eq:interleave_metric}
    \Delta_{AB} = \frac{wins(A) + 0.5 * ties(A, B)}{wins(A) + wins(B) + ties(A,B)} - 0.5,
\end{equation}
where $wins(A)$ (or $wins(B)$) is a counter that would be increased by 1 if the results produced by $A$ (or $B$) is preferred by the user, and $ties(A, B)$ is increased by 1 otherwise. Intuitively, $\Delta_{AB} > 0$ means $A$ is better than $B$. 
To further reduce the bias in the evaluation, we further introduce a time-weighted version of $\Delta_{AB}$, denoted as $\Delta_{AB}\text{-tw}$, where the counters are weighted by the post-click dwell time (mapped to $[0, 1]$ with a sigmoid function). As such, the $\Delta_{AB}\text{-tw}$ can better reflect the relevance of the results with higher confidence, as users would stay in a web page longer if it is relevant.

We conduct balanced interleaving experiments~\cite{chuklin2015comparative} for comparing the ERNIE-enhanced system against the old retrieval system. The results are shown in Table \ref{exp:interleaving}, which comprise the performance of different modules and for different types of queries. 
\begin{itemize}[leftmargin=5mm]
    \item First, we observe $\Delta_{AB} > 0$ and $\Delta_{AB}\text{-tw} > 0$ in all the experiments, which indicate that the new system can increase user clicks with more relevant web documents. 
    \item Second, the results also indicate that both the ERNIE-based retrieval and post-retrieval filtering can boost the effectiveness of the system, while the ERNIE-based retrieval generally achieves larger gain than the post-retrieval filtering, as it is effected before the filtering stage.
    \item Third, we can see that the new system can achieve better performance on the queries with low search frequency, i.e., tail query. This validates that our proposed retrieval system has more significant improvement for the tail queries. 
    \item Also, the improvements on long queries (e.g., natural language queries) is larger than on short queries, which might indicate that the system can better handle the natural language queries. 
\end{itemize}

\begin{table}[t]
\caption{Relative improvement on manual evaluation.} \label{exp:manual}
\vspace{-2mm}
\scalebox{+0.93}{
\begin{tabular}{l|cc|cc}
\hline \hline
\multirow{2}{*}{} & \multicolumn{2}{c|}{ERNIE-based retrieval} & \multicolumn{2}{c}{Post-retrieval filtering} \\ \cline{2-5} 
                  & Rand                 & Tail                & Rand                  & Tail                  \\ \hline
$\Delta_{DCG}$    & +0.17\%              & +0.22\%             & +0.10\%               & +0.65\%               \\ \hline
$\Delta_{GSB}$    & +3.50\%              & +7.50\%             & +3.96\%               & +3.13\%               \\ \hline \hline
\multicolumn{5}{c}{*All the values are statistically significant ($t$-test with $p<0.05$).}
\end{tabular}}
\vspace{-2mm}
\end{table}

\subsection{Online A/B Test}
We also conduct online A/B Test that compares the new system with the old system for one week. In the online A/B test, We mainly focus on the metrics that directly related to user experience. The results show that the proposed retrieval system can largely improve the overall user experience of the search engine. In particular, the number of click behaviors has increased by 0.31\%. The number of click-and-stay behavior has increased by 0.57\%. The average post-click dwell time has increased by 0.62\%. The click-through rate has increased by 0.3\%. All the reported values are statistically significant with $p<0.05$.  This shows that accurate semantic modeling and semantic matching in the retrieval stage is very helpful for improving the user engagement for the search engine.

\subsection{Manual Evaluation for Online Cases}
To more comprehensively show the impact of our proposed system, we further conduct a manual evaluation on the final ranking results with some real user-generated queries. This directly reflects the quality of the results exposed to the end users. 

\noindent\textbf{Data preparation}.
We log a set of (several hundreds) queries and the corresponding final impressions, i.e., the top-ranked web documents in the final ranking stage, using the ERNIE-enhanced and the old retrieval systems. Note that the data logging is conducted by multiple rounds to eliminate randomness. We filter out whose queries that have identical impressions between the two systems, and then use the rest for the manual evaluation.
The relative improvement validated by manual evaluation is given in Table \ref{exp:manual}.

\noindent\textbf{Discounted Cumulative Gain (DCG)}. We first log a dataset and manually label the data with 0 to 4 grades, and then report the relative improvement w.r.t. the average Discounted Cumulative Gain (DCG) over the top-4 final results of all queries). The DCG is a widely-used metric and thus we omit its definition here. As shown in Figure \ref{exp:manual}, the results again show that our proposed system is able to improve the effectiveness of retrieval, especially for tail queries.

\noindent\textbf{Side-by-side comparison}. Besides, we also conduct a side-by-side comparison between the two systems. We log another dataset, and require the human experts to judge whether the new system or the base system gives better final results. Here, the relative gain is measured Good vs. Same vs. Bad (GSB) as
\begin{equation}
    \Delta_{GSB} = \frac{\# \text{Good} - \# \text{Bad}}{\# \text{Good} + \# \text{Same} + \# \text{Bad}},
\end{equation}
where $\# \text{Good}$ (or \# \text{Bad}) indicates the number of queries that the new system provides better (or worse) final results.
Table \ref{exp:manual} shows that for both random queries and tail queries, the new ERNIE-enhanced system can significantly outperform the base system.

\section{Conclusion}
In this paper, we described the novel retrieval system that is facilitated by pretrained language model (i.e., ERNIE). We developed and deployed the system in Baidu Search, which is highly effective in conducting semantic retrieval for web search. The system employs 1) an ERNIE-based retrieval model, 2) a multi-stage training paradigm and 3) a unified workflow for the retrieval system. Extensive offline and online experiments has shown that the retrieval system can significantly improve the effectiveness and general usability of the search engine. 

\clearpage
\bibliographystyle{ACM-Reference-Format}
\bibliography{reference} 


\begin{thebibliography}{53}


\ifx \showCODEN    \undefined \def \showCODEN     #1{\unskip}     \fi
\ifx \showDOI      \undefined \def \showDOI       #1{#1}\fi
\ifx \showISBNx    \undefined \def \showISBNx     #1{\unskip}     \fi
\ifx \showISBNxiii \undefined \def \showISBNxiii  #1{\unskip}     \fi
\ifx \showISSN     \undefined \def \showISSN      #1{\unskip}     \fi
\ifx \showLCCN     \undefined \def \showLCCN      #1{\unskip}     \fi
\ifx \shownote     \undefined \def \shownote      #1{#1}          \fi
\ifx \showarticletitle \undefined \def \showarticletitle #1{#1}   \fi
\ifx \showURL      \undefined \def \showURL       {\relax}        \fi
\providecommand\bibfield[2]{#2}
\providecommand\bibinfo[2]{#2}
\providecommand\natexlab[1]{#1}
\providecommand\showeprint[2][]{arXiv:#2}

\bibitem[\protect\citeauthoryear{Cai, Fan, Guo, Sun, Zhang, and Cheng}{Cai
  et~al\mbox{.}}{2021}]%
        {cai2021semantic}
\bibfield{author}{\bibinfo{person}{Yinqiong Cai}, \bibinfo{person}{Yixing Fan},
  \bibinfo{person}{Jiafeng Guo}, \bibinfo{person}{Fei Sun},
  \bibinfo{person}{Ruqing Zhang}, {and} \bibinfo{person}{Xueqi Cheng}.}
  \bibinfo{year}{2021}\natexlab{}.
\newblock \showarticletitle{Semantic Models for the First-stage Retrieval: A
  Comprehensive Review}.
\newblock \bibinfo{journal}{\emph{arXiv preprint arXiv:2103.04831}}
  (\bibinfo{year}{2021}).
\newblock


\bibitem[\protect\citeauthoryear{Chang, Yu, Chang, Yang, and Kumar}{Chang
  et~al\mbox{.}}{2020}]%
        {chang2020pre}
\bibfield{author}{\bibinfo{person}{Wei-Cheng Chang}, \bibinfo{person}{Felix~X
  Yu}, \bibinfo{person}{Yin-Wen Chang}, \bibinfo{person}{Yiming Yang}, {and}
  \bibinfo{person}{Sanjiv Kumar}.} \bibinfo{year}{2020}\natexlab{}.
\newblock \showarticletitle{Pre-training tasks for embedding-based large-scale
  retrieval}.
\newblock \bibinfo{journal}{\emph{arXiv preprint arXiv:2002.03932}}
  (\bibinfo{year}{2020}).
\newblock


\bibitem[\protect\citeauthoryear{Chuklin, Schuth, Zhou, and Rijke}{Chuklin
  et~al\mbox{.}}{2015}]%
        {chuklin2015comparative}
\bibfield{author}{\bibinfo{person}{Aleksandr Chuklin}, \bibinfo{person}{Anne
  Schuth}, \bibinfo{person}{Ke Zhou}, {and} \bibinfo{person}{Maarten~De
  Rijke}.} \bibinfo{year}{2015}\natexlab{}.
\newblock \showarticletitle{A comparative analysis of interleaving methods for
  aggregated search}.
\newblock \bibinfo{journal}{\emph{TOIS}} \bibinfo{volume}{33},
  \bibinfo{number}{2} (\bibinfo{year}{2015}), \bibinfo{pages}{1--38}.
\newblock


\bibitem[\protect\citeauthoryear{Devlin, Chang, Lee, and Toutanova}{Devlin
  et~al\mbox{.}}{2018}]%
        {devlin2018bert}
\bibfield{author}{\bibinfo{person}{Jacob Devlin}, \bibinfo{person}{Ming-Wei
  Chang}, \bibinfo{person}{Kenton Lee}, {and} \bibinfo{person}{Kristina
  Toutanova}.} \bibinfo{year}{2018}\natexlab{}.
\newblock \showarticletitle{Bert: Pre-training of deep bidirectional
  transformers for language understanding}.
\newblock \bibinfo{journal}{\emph{arXiv preprint arXiv:1810.04805}}
  (\bibinfo{year}{2018}).
\newblock


\bibitem[\protect\citeauthoryear{Dong and Shen}{Dong and Shen}{2018}]%
        {dong2018triplet}
\bibfield{author}{\bibinfo{person}{Xingping Dong} {and}
  \bibinfo{person}{Jianbing Shen}.} \bibinfo{year}{2018}\natexlab{}.
\newblock \showarticletitle{Triplet loss in siamese network for object
  tracking}. In \bibinfo{booktitle}{\emph{ECCV}}. \bibinfo{pages}{459--474}.
\newblock


\bibitem[\protect\citeauthoryear{Firat, Sankaran, Al-Onaizan, Vural, and
  Cho}{Firat et~al\mbox{.}}{2016}]%
        {firat2016zero}
\bibfield{author}{\bibinfo{person}{Orhan Firat}, \bibinfo{person}{Baskaran
  Sankaran}, \bibinfo{person}{Yaser Al-Onaizan}, \bibinfo{person}{Fatos
  T~Yarman Vural}, {and} \bibinfo{person}{Kyunghyun Cho}.}
  \bibinfo{year}{2016}\natexlab{}.
\newblock \showarticletitle{Zero-Resource Translation with Multi-Lingual Neural
  Machine Translation}. In \bibinfo{booktitle}{\emph{EMNLP}}.
  \bibinfo{pages}{268--277}.
\newblock


\bibitem[\protect\citeauthoryear{Gao, Pantel, Gamon, He, and Deng}{Gao
  et~al\mbox{.}}{2014}]%
        {gao2019modeling}
\bibfield{author}{\bibinfo{person}{Jianfeng Gao}, \bibinfo{person}{Patrick
  Pantel}, \bibinfo{person}{Michael Gamon}, \bibinfo{person}{Xiaodong He},
  {and} \bibinfo{person}{Li Deng}.} \bibinfo{year}{2014}\natexlab{}.
\newblock \showarticletitle{Modeling interestingness with deep neural
  networks}.
\newblock  (\bibinfo{year}{2014}).
\newblock


\bibitem[\protect\citeauthoryear{Gao, Toutanova, and Yih}{Gao
  et~al\mbox{.}}{2011}]%
        {gao2011clickthrough}
\bibfield{author}{\bibinfo{person}{Jianfeng Gao}, \bibinfo{person}{Kristina
  Toutanova}, {and} \bibinfo{person}{Wen-tau Yih}.}
  \bibinfo{year}{2011}\natexlab{}.
\newblock \showarticletitle{Clickthrough-based latent semantic models for web
  search}. In \bibinfo{booktitle}{\emph{SIGIR}}. \bibinfo{pages}{675--684}.
\newblock


\bibitem[\protect\citeauthoryear{Gu, Ding, Wang, Zou, Liu, and Yin}{Gu
  et~al\mbox{.}}{2020}]%
        {gu2020deep}
\bibfield{author}{\bibinfo{person}{Yulong Gu}, \bibinfo{person}{Zhuoye Ding},
  \bibinfo{person}{Shuaiqiang Wang}, \bibinfo{person}{Lixin Zou},
  \bibinfo{person}{Yiding Liu}, {and} \bibinfo{person}{Dawei Yin}.}
  \bibinfo{year}{2020}\natexlab{}.
\newblock \showarticletitle{Deep Multifaceted Transformers for Multi-objective
  Ranking in Large-Scale E-commerce Recommender Systems}. In
  \bibinfo{booktitle}{\emph{CIKM}}. \bibinfo{pages}{2493--2500}.
\newblock


\bibitem[\protect\citeauthoryear{Guo, Fan, Ai, and Croft}{Guo
  et~al\mbox{.}}{2016}]%
        {guo2016deep}
\bibfield{author}{\bibinfo{person}{Jiafeng Guo}, \bibinfo{person}{Yixing Fan},
  \bibinfo{person}{Qingyao Ai}, {and} \bibinfo{person}{W~Bruce Croft}.}
  \bibinfo{year}{2016}\natexlab{}.
\newblock \showarticletitle{A deep relevance matching model for ad-hoc
  retrieval}. In \bibinfo{booktitle}{\emph{CIKM}}. \bibinfo{pages}{55--64}.
\newblock


\bibitem[\protect\citeauthoryear{Haldar, Ramanathan, Sax, Abdool, Zhang,
  Mansawala, Yang, Turnbull, and Liao}{Haldar et~al\mbox{.}}{2020}]%
        {haldar2020improving}
\bibfield{author}{\bibinfo{person}{Malay Haldar}, \bibinfo{person}{Prashant
  Ramanathan}, \bibinfo{person}{Tyler Sax}, \bibinfo{person}{Mustafa Abdool},
  \bibinfo{person}{Lanbo Zhang}, \bibinfo{person}{Aamir Mansawala},
  \bibinfo{person}{Shulin Yang}, \bibinfo{person}{Bradley Turnbull}, {and}
  \bibinfo{person}{Junshuo Liao}.} \bibinfo{year}{2020}\natexlab{}.
\newblock \showarticletitle{Improving Deep Learning For Airbnb Search}. In
  \bibinfo{booktitle}{\emph{KDD}}. \bibinfo{pages}{2822--2830}.
\newblock


\bibitem[\protect\citeauthoryear{Hinton, Vinyals, and Dean}{Hinton
  et~al\mbox{.}}{2015}]%
        {hinton2015distilling}
\bibfield{author}{\bibinfo{person}{Geoffrey Hinton}, \bibinfo{person}{Oriol
  Vinyals}, {and} \bibinfo{person}{Jeff Dean}.}
  \bibinfo{year}{2015}\natexlab{}.
\newblock \showarticletitle{Distilling the knowledge in a neural network}.
\newblock \bibinfo{journal}{\emph{arXiv preprint arXiv:1503.02531}}
  (\bibinfo{year}{2015}).
\newblock


\bibitem[\protect\citeauthoryear{Huang, Sharma, Sun, Xia, Zhang, Pronin,
  Padmanabhan, Ottaviano, and Yang}{Huang et~al\mbox{.}}{2020}]%
        {huang2020embedding}
\bibfield{author}{\bibinfo{person}{Jui-Ting Huang}, \bibinfo{person}{Ashish
  Sharma}, \bibinfo{person}{Shuying Sun}, \bibinfo{person}{Li Xia},
  \bibinfo{person}{David Zhang}, \bibinfo{person}{Philip Pronin},
  \bibinfo{person}{Janani Padmanabhan}, \bibinfo{person}{Giuseppe Ottaviano},
  {and} \bibinfo{person}{Linjun Yang}.} \bibinfo{year}{2020}\natexlab{}.
\newblock \showarticletitle{Embedding-based retrieval in facebook search}. In
  \bibinfo{booktitle}{\emph{KDD}}. \bibinfo{pages}{2553--2561}.
\newblock


\bibitem[\protect\citeauthoryear{Huang, He, Gao, Deng, Acero, and Heck}{Huang
  et~al\mbox{.}}{2013}]%
        {huang2013learning}
\bibfield{author}{\bibinfo{person}{Po-Sen Huang}, \bibinfo{person}{Xiaodong
  He}, \bibinfo{person}{Jianfeng Gao}, \bibinfo{person}{Li Deng},
  \bibinfo{person}{Alex Acero}, {and} \bibinfo{person}{Larry Heck}.}
  \bibinfo{year}{2013}\natexlab{}.
\newblock \showarticletitle{Learning deep structured semantic models for web
  search using clickthrough data}. In \bibinfo{booktitle}{\emph{CIKM}}.
  \bibinfo{pages}{2333--2338}.
\newblock


\bibitem[\protect\citeauthoryear{Hubara, Courbariaux, Soudry, El-Yaniv, and
  Bengio}{Hubara et~al\mbox{.}}{2017}]%
        {hubara2017quantized}
\bibfield{author}{\bibinfo{person}{Itay Hubara}, \bibinfo{person}{Matthieu
  Courbariaux}, \bibinfo{person}{Daniel Soudry}, \bibinfo{person}{Ran
  El-Yaniv}, {and} \bibinfo{person}{Yoshua Bengio}.}
  \bibinfo{year}{2017}\natexlab{}.
\newblock \showarticletitle{Quantized neural networks: Training neural networks
  with low precision weights and activations}.
\newblock \bibinfo{journal}{\emph{JMLR}} \bibinfo{volume}{18},
  \bibinfo{number}{1} (\bibinfo{year}{2017}), \bibinfo{pages}{6869--6898}.
\newblock


\bibitem[\protect\citeauthoryear{Humeau, Shuster, Lachaux, and Weston}{Humeau
  et~al\mbox{.}}{2019}]%
        {humeau2019poly}
\bibfield{author}{\bibinfo{person}{Samuel Humeau}, \bibinfo{person}{Kurt
  Shuster}, \bibinfo{person}{Marie-Anne Lachaux}, {and} \bibinfo{person}{Jason
  Weston}.} \bibinfo{year}{2019}\natexlab{}.
\newblock \showarticletitle{Poly-encoders: Transformer architectures and
  pre-training strategies for fast and accurate multi-sentence scoring}.
\newblock \bibinfo{journal}{\emph{arXiv preprint arXiv:1905.01969}}
  (\bibinfo{year}{2019}).
\newblock


\bibitem[\protect\citeauthoryear{Johnson, Schuster, Le, Krikun, Wu, Chen,
  Thorat, Vi{\'e}gas, Wattenberg, Corrado, et~al\mbox{.}}{Johnson
  et~al\mbox{.}}{2017}]%
        {johnson2017google}
\bibfield{author}{\bibinfo{person}{Melvin Johnson}, \bibinfo{person}{Mike
  Schuster}, \bibinfo{person}{Quoc Le}, \bibinfo{person}{Maxim Krikun},
  \bibinfo{person}{Yonghui Wu}, \bibinfo{person}{Zhifeng Chen},
  \bibinfo{person}{Nikhil Thorat}, \bibinfo{person}{Fernanda Vi{\'e}gas},
  \bibinfo{person}{Martin Wattenberg}, \bibinfo{person}{Greg Corrado},
  {et~al\mbox{.}}} \bibinfo{year}{2017}\natexlab{}.
\newblock \showarticletitle{Google’s Multilingual Neural Machine Translation
  System: Enabling Zero-Shot Translation}.
\newblock \bibinfo{journal}{\emph{TACL}}  \bibinfo{volume}{5}
  (\bibinfo{year}{2017}), \bibinfo{pages}{339--351}.
\newblock


\bibitem[\protect\citeauthoryear{Karpukhin, Oguz, Min, Lewis, Wu, Edunov, Chen,
  and Yih}{Karpukhin et~al\mbox{.}}{2020}]%
        {karpukhin2020dense}
\bibfield{author}{\bibinfo{person}{Vladimir Karpukhin}, \bibinfo{person}{Barlas
  Oguz}, \bibinfo{person}{Sewon Min}, \bibinfo{person}{Patrick Lewis},
  \bibinfo{person}{Ledell Wu}, \bibinfo{person}{Sergey Edunov},
  \bibinfo{person}{Danqi Chen}, {and} \bibinfo{person}{Wen-tau Yih}.}
  \bibinfo{year}{2020}\natexlab{}.
\newblock \showarticletitle{Dense Passage Retrieval for Open-Domain Question
  Answering}. In \bibinfo{booktitle}{\emph{Proceedings of the 2020 Conference
  on Empirical Methods in Natural Language Processing (EMNLP)}}.
  \bibinfo{pages}{6769--6781}.
\newblock


\bibitem[\protect\citeauthoryear{Lee, Chang, and Toutanova}{Lee
  et~al\mbox{.}}{2019}]%
        {lee2019latent}
\bibfield{author}{\bibinfo{person}{Kenton Lee}, \bibinfo{person}{Ming-Wei
  Chang}, {and} \bibinfo{person}{Kristina Toutanova}.}
  \bibinfo{year}{2019}\natexlab{}.
\newblock \showarticletitle{Latent Retrieval for Weakly Supervised Open Domain
  Question Answering}. In \bibinfo{booktitle}{\emph{ACL}}.
  \bibinfo{pages}{6086--6096}.
\newblock


\bibitem[\protect\citeauthoryear{Li and Xu}{Li and Xu}{2014}]%
        {li2014semantic}
\bibfield{author}{\bibinfo{person}{Hang Li} {and} \bibinfo{person}{Jun Xu}.}
  \bibinfo{year}{2014}\natexlab{}.
\newblock \showarticletitle{Semantic matching in search}.
\newblock \bibinfo{journal}{\emph{Foundations and Trends in Information
  retrieval}} \bibinfo{volume}{7}, \bibinfo{number}{5} (\bibinfo{year}{2014}),
  \bibinfo{pages}{343--469}.
\newblock


\bibitem[\protect\citeauthoryear{Liu, Gu, Ding, Gao, Guo, Bao, and Yan}{Liu
  et~al\mbox{.}}{2020}]%
        {liu2020decoupled}
\bibfield{author}{\bibinfo{person}{Yiding Liu}, \bibinfo{person}{Yulong Gu},
  \bibinfo{person}{Zhuoye Ding}, \bibinfo{person}{Junchao Gao},
  \bibinfo{person}{Ziyi Guo}, \bibinfo{person}{Yongjun Bao}, {and}
  \bibinfo{person}{Weipeng Yan}.} \bibinfo{year}{2020}\natexlab{}.
\newblock \showarticletitle{Decoupled Graph Convolution Network for Inferring
  Substitutable and Complementary Items}. In \bibinfo{booktitle}{\emph{CIKM}}.
  \bibinfo{pages}{2621--2628}.
\newblock


\bibitem[\protect\citeauthoryear{Liu, Ott, Goyal, Du, Joshi, Chen, Levy, Lewis,
  Zettlemoyer, and Stoyanov}{Liu et~al\mbox{.}}{2019}]%
        {liu2019roberta}
\bibfield{author}{\bibinfo{person}{Yinhan Liu}, \bibinfo{person}{Myle Ott},
  \bibinfo{person}{Naman Goyal}, \bibinfo{person}{Jingfei Du},
  \bibinfo{person}{Mandar Joshi}, \bibinfo{person}{Danqi Chen},
  \bibinfo{person}{Omer Levy}, \bibinfo{person}{Mike Lewis},
  \bibinfo{person}{Luke Zettlemoyer}, {and} \bibinfo{person}{Veselin
  Stoyanov}.} \bibinfo{year}{2019}\natexlab{}.
\newblock \showarticletitle{Roberta: A robustly optimized bert pretraining
  approach}.
\newblock \bibinfo{journal}{\emph{arXiv preprint arXiv:1907.11692}}
  (\bibinfo{year}{2019}).
\newblock


\bibitem[\protect\citeauthoryear{Lu, Abrego, Ma, Ni, and Yang}{Lu
  et~al\mbox{.}}{2020a}]%
        {lu2020neural}
\bibfield{author}{\bibinfo{person}{Jing Lu}, \bibinfo{person}{Gustavo~Hernandez
  Abrego}, \bibinfo{person}{Ji Ma}, \bibinfo{person}{Jianmo Ni}, {and}
  \bibinfo{person}{Yinfei Yang}.} \bibinfo{year}{2020}\natexlab{a}.
\newblock \showarticletitle{Neural Passage Retrieval with Improved Negative
  Contrast}.
\newblock \bibinfo{journal}{\emph{arXiv preprint arXiv:2010.12523}}
  (\bibinfo{year}{2020}).
\newblock


\bibitem[\protect\citeauthoryear{Lu, Jiao, and Zhang}{Lu
  et~al\mbox{.}}{2020b}]%
        {lu2020twinbert}
\bibfield{author}{\bibinfo{person}{Wenhao Lu}, \bibinfo{person}{Jian Jiao},
  {and} \bibinfo{person}{Ruofei Zhang}.} \bibinfo{year}{2020}\natexlab{b}.
\newblock \showarticletitle{TwinBERT: Distilling Knowledge to Twin-Structured
  Compressed BERT Models for Large-Scale Retrieval}. In
  \bibinfo{booktitle}{\emph{CIKM}}. \bibinfo{pages}{2645--2652}.
\newblock


\bibitem[\protect\citeauthoryear{Lu and Li}{Lu and Li}{2013}]%
        {lu2013deep}
\bibfield{author}{\bibinfo{person}{Zhengdong Lu} {and} \bibinfo{person}{Hang
  Li}.} \bibinfo{year}{2013}\natexlab{}.
\newblock \showarticletitle{A deep architecture for matching short texts}.
\newblock \bibinfo{journal}{\emph{ç}}  \bibinfo{volume}{26}
  (\bibinfo{year}{2013}), \bibinfo{pages}{1367--1375}.
\newblock


\bibitem[\protect\citeauthoryear{Luan, Eisenstein, Toutanova, and Collins}{Luan
  et~al\mbox{.}}{2020}]%
        {luan2020sparse}
\bibfield{author}{\bibinfo{person}{Yi Luan}, \bibinfo{person}{Jacob
  Eisenstein}, \bibinfo{person}{Kristina Toutanova}, {and}
  \bibinfo{person}{Michael Collins}.} \bibinfo{year}{2020}\natexlab{}.
\newblock \showarticletitle{Sparse, dense, and attentional representations for
  text retrieval}.
\newblock \bibinfo{journal}{\emph{arXiv preprint arXiv:2005.00181}}
  (\bibinfo{year}{2020}).
\newblock


\bibitem[\protect\citeauthoryear{Mitra, Craswell, et~al\mbox{.}}{Mitra
  et~al\mbox{.}}{2018}]%
        {mitra2018introduction}
\bibfield{author}{\bibinfo{person}{Bhaskar Mitra}, \bibinfo{person}{Nick
  Craswell}, {et~al\mbox{.}}} \bibinfo{year}{2018}\natexlab{}.
\newblock \bibinfo{booktitle}{\emph{An introduction to neural information
  retrieval}}.
\newblock \bibinfo{publisher}{Now Foundations and Trends}.
\newblock


\bibitem[\protect\citeauthoryear{Mitra, Diaz, and Craswell}{Mitra
  et~al\mbox{.}}{2017}]%
        {mitra2017learning}
\bibfield{author}{\bibinfo{person}{Bhaskar Mitra}, \bibinfo{person}{Fernando
  Diaz}, {and} \bibinfo{person}{Nick Craswell}.}
  \bibinfo{year}{2017}\natexlab{}.
\newblock \showarticletitle{Learning to match using local and distributed
  representations of text for web search}. In \bibinfo{booktitle}{\emph{WWW}}.
  \bibinfo{pages}{1291--1299}.
\newblock


\bibitem[\protect\citeauthoryear{Palangi, Deng, Shen, Gao, He, Chen, Song, and
  Ward}{Palangi et~al\mbox{.}}{2014}]%
        {palangi2014semantic}
\bibfield{author}{\bibinfo{person}{Hamid Palangi}, \bibinfo{person}{Li Deng},
  \bibinfo{person}{Yelong Shen}, \bibinfo{person}{Jianfeng Gao},
  \bibinfo{person}{Xiaodong He}, \bibinfo{person}{Jianshu Chen},
  \bibinfo{person}{Xinying Song}, {and} \bibinfo{person}{R Ward}.}
  \bibinfo{year}{2014}\natexlab{}.
\newblock \showarticletitle{Semantic modelling with long-short-term memory for
  information retrieval}.
\newblock \bibinfo{journal}{\emph{arXiv preprint arXiv:1412.6629}}
  (\bibinfo{year}{2014}).
\newblock


\bibitem[\protect\citeauthoryear{Palangi, Palangi, Deng, Shen, Gao, He, Chen,
  Song, and Ward}{Palangi et~al\mbox{.}}{2015}]%
        {palangi2015deep}
\bibfield{author}{\bibinfo{person}{Hamid Palangi}, \bibinfo{person}{H Palangi},
  \bibinfo{person}{L Deng}, \bibinfo{person}{Y Shen}, \bibinfo{person}{J Gao},
  \bibinfo{person}{X He}, \bibinfo{person}{J Chen}, \bibinfo{person}{X Song},
  {and} \bibinfo{person}{R Ward}.} \bibinfo{year}{2015}\natexlab{}.
\newblock \bibinfo{title}{Deep Sentence Embedding Using the Long Short Term
  Memory Network: Analysis and Application to Information Retrieval. arXiv.
  org}.
\newblock
\newblock


\bibitem[\protect\citeauthoryear{Peters, Neumann, Iyyer, Gardner, Clark, Lee,
  and Zettlemoyer}{Peters et~al\mbox{.}}{2018}]%
        {peters2018deep}
\bibfield{author}{\bibinfo{person}{Matthew~E Peters}, \bibinfo{person}{Mark
  Neumann}, \bibinfo{person}{Mohit Iyyer}, \bibinfo{person}{Matt Gardner},
  \bibinfo{person}{Christopher Clark}, \bibinfo{person}{Kenton Lee}, {and}
  \bibinfo{person}{Luke Zettlemoyer}.} \bibinfo{year}{2018}\natexlab{}.
\newblock \showarticletitle{Deep contextualized word representations}.
\newblock \bibinfo{journal}{\emph{arXiv preprint arXiv:1802.05365}}
  (\bibinfo{year}{2018}).
\newblock


\bibitem[\protect\citeauthoryear{Pruksachatkun, Phang, Liu, Htut, Zhang, Pang,
  Vania, Kann, and Bowman}{Pruksachatkun et~al\mbox{.}}{2020}]%
        {pruksachatkun2020intermediate}
\bibfield{author}{\bibinfo{person}{Yada Pruksachatkun}, \bibinfo{person}{Jason
  Phang}, \bibinfo{person}{Haokun Liu}, \bibinfo{person}{Phu~Mon Htut},
  \bibinfo{person}{Xiaoyi Zhang}, \bibinfo{person}{Richard~Yuanzhe Pang},
  \bibinfo{person}{Clara Vania}, \bibinfo{person}{Katharina Kann}, {and}
  \bibinfo{person}{Samuel Bowman}.} \bibinfo{year}{2020}\natexlab{}.
\newblock \showarticletitle{Intermediate-Task Transfer Learning with Pretrained
  Language Models: When and Why Does It Work?}. In
  \bibinfo{booktitle}{\emph{ACL}}. \bibinfo{pages}{5231--5247}.
\newblock


\bibitem[\protect\citeauthoryear{Salakhutdinov and Hinton}{Salakhutdinov and
  Hinton}{2009}]%
        {salakhutdinov2009semantic}
\bibfield{author}{\bibinfo{person}{Ruslan Salakhutdinov} {and}
  \bibinfo{person}{Geoffrey Hinton}.} \bibinfo{year}{2009}\natexlab{}.
\newblock \showarticletitle{Semantic hashing}.
\newblock \bibinfo{journal}{\emph{IJAR}} \bibinfo{volume}{50},
  \bibinfo{number}{7} (\bibinfo{year}{2009}), \bibinfo{pages}{969--978}.
\newblock


\bibitem[\protect\citeauthoryear{Sch{\"u}tze, Manning, and
  Raghavan}{Sch{\"u}tze et~al\mbox{.}}{2008}]%
        {schutze2008introduction}
\bibfield{author}{\bibinfo{person}{Hinrich Sch{\"u}tze},
  \bibinfo{person}{Christopher~D Manning}, {and} \bibinfo{person}{Prabhakar
  Raghavan}.} \bibinfo{year}{2008}\natexlab{}.
\newblock \bibinfo{booktitle}{\emph{Introduction to information retrieval}}.
  Vol.~\bibinfo{volume}{39}.
\newblock \bibinfo{publisher}{Cambridge University Press Cambridge}.
\newblock


\bibitem[\protect\citeauthoryear{Severyn and Moschitti}{Severyn and
  Moschitti}{2015}]%
        {severyn2015learning}
\bibfield{author}{\bibinfo{person}{Aliaksei Severyn} {and}
  \bibinfo{person}{Alessandro Moschitti}.} \bibinfo{year}{2015}\natexlab{}.
\newblock \showarticletitle{Learning to rank short text pairs with
  convolutional deep neural networks}. In \bibinfo{booktitle}{\emph{SIGIR}}.
  \bibinfo{pages}{373--382}.
\newblock


\bibitem[\protect\citeauthoryear{Shen, He, Gao, Deng, and Mesnil}{Shen
  et~al\mbox{.}}{2014a}]%
        {shen2014latent}
\bibfield{author}{\bibinfo{person}{Yelong Shen}, \bibinfo{person}{Xiaodong He},
  \bibinfo{person}{Jianfeng Gao}, \bibinfo{person}{Li Deng}, {and}
  \bibinfo{person}{Gr{\'e}goire Mesnil}.} \bibinfo{year}{2014}\natexlab{a}.
\newblock \showarticletitle{A latent semantic model with convolutional-pooling
  structure for information retrieval}. In \bibinfo{booktitle}{\emph{CIKM}}.
  \bibinfo{pages}{101--110}.
\newblock


\bibitem[\protect\citeauthoryear{Shen, He, Gao, Deng, and Mesnil}{Shen
  et~al\mbox{.}}{2014b}]%
        {shen2014learning}
\bibfield{author}{\bibinfo{person}{Yelong Shen}, \bibinfo{person}{Xiaodong He},
  \bibinfo{person}{Jianfeng Gao}, \bibinfo{person}{Li Deng}, {and}
  \bibinfo{person}{Gr{\'e}goire Mesnil}.} \bibinfo{year}{2014}\natexlab{b}.
\newblock \showarticletitle{Learning semantic representations using
  convolutional neural networks for web search}. In
  \bibinfo{booktitle}{\emph{WWW}}. \bibinfo{pages}{373--374}.
\newblock


\bibitem[\protect\citeauthoryear{Sun, Wang, Li, Feng, Chen, Zhang, Tian, Zhu,
  Tian, and Wu}{Sun et~al\mbox{.}}{2019}]%
        {sun2019ernie}
\bibfield{author}{\bibinfo{person}{Yu Sun}, \bibinfo{person}{Shuohuan Wang},
  \bibinfo{person}{Yukun Li}, \bibinfo{person}{Shikun Feng},
  \bibinfo{person}{Xuyi Chen}, \bibinfo{person}{Han Zhang},
  \bibinfo{person}{Xin Tian}, \bibinfo{person}{Danxiang Zhu},
  \bibinfo{person}{Hao Tian}, {and} \bibinfo{person}{Hua Wu}.}
  \bibinfo{year}{2019}\natexlab{}.
\newblock \showarticletitle{Ernie: Enhanced representation through knowledge
  integration}.
\newblock \bibinfo{journal}{\emph{arXiv preprint arXiv:1904.09223}}
  (\bibinfo{year}{2019}).
\newblock


\bibitem[\protect\citeauthoryear{Vaswani, Shazeer, Parmar, Uszkoreit, Jones,
  Gomez, Kaiser, and Polosukhin}{Vaswani et~al\mbox{.}}{2017}]%
        {vaswani2017attention}
\bibfield{author}{\bibinfo{person}{Ashish Vaswani}, \bibinfo{person}{Noam
  Shazeer}, \bibinfo{person}{Niki Parmar}, \bibinfo{person}{Jakob Uszkoreit},
  \bibinfo{person}{Llion Jones}, \bibinfo{person}{Aidan~N Gomez},
  \bibinfo{person}{{\L}ukasz Kaiser}, {and} \bibinfo{person}{Illia
  Polosukhin}.} \bibinfo{year}{2017}\natexlab{}.
\newblock \showarticletitle{Attention is all you need}. In
  \bibinfo{booktitle}{\emph{NeurIPS}}. \bibinfo{pages}{5998--6008}.
\newblock


\bibitem[\protect\citeauthoryear{Wan, Lan, Guo, Xu, Pang, and Cheng}{Wan
  et~al\mbox{.}}{2016}]%
        {wan2016deep}
\bibfield{author}{\bibinfo{person}{Shengxian Wan}, \bibinfo{person}{Yanyan
  Lan}, \bibinfo{person}{Jiafeng Guo}, \bibinfo{person}{Jun Xu},
  \bibinfo{person}{Liang Pang}, {and} \bibinfo{person}{Xueqi Cheng}.}
  \bibinfo{year}{2016}\natexlab{}.
\newblock \showarticletitle{A deep architecture for semantic matching with
  multiple positional sentence representations}. In
  \bibinfo{booktitle}{\emph{AAAI}}, Vol.~\bibinfo{volume}{30}.
\newblock


\bibitem[\protect\citeauthoryear{Xia, Tan, Tian, Qin, Yu, and Liu}{Xia
  et~al\mbox{.}}{2018}]%
        {xia2018model}
\bibfield{author}{\bibinfo{person}{Yingce Xia}, \bibinfo{person}{Xu Tan},
  \bibinfo{person}{Fei Tian}, \bibinfo{person}{Tao Qin},
  \bibinfo{person}{Nenghai Yu}, {and} \bibinfo{person}{Tie-Yan Liu}.}
  \bibinfo{year}{2018}\natexlab{}.
\newblock \showarticletitle{Model-level dual learning}. In
  \bibinfo{booktitle}{\emph{International Conference on Machine Learning}}.
  PMLR, \bibinfo{pages}{5383--5392}.
\newblock


\bibitem[\protect\citeauthoryear{Xiong, Xiong, Li, Tang, Liu, Bennett, Ahmed,
  and Overwijk}{Xiong et~al\mbox{.}}{2020}]%
        {xiong2020approximate}
\bibfield{author}{\bibinfo{person}{Lee Xiong}, \bibinfo{person}{Chenyan Xiong},
  \bibinfo{person}{Ye Li}, \bibinfo{person}{Kwok-Fung Tang},
  \bibinfo{person}{Jialin Liu}, \bibinfo{person}{Paul Bennett},
  \bibinfo{person}{Junaid Ahmed}, {and} \bibinfo{person}{Arnold Overwijk}.}
  \bibinfo{year}{2020}\natexlab{}.
\newblock \showarticletitle{Approximate nearest neighbor negative contrastive
  learning for dense text retrieval}.
\newblock \bibinfo{journal}{\emph{arXiv preprint arXiv:2007.00808}}
  (\bibinfo{year}{2020}).
\newblock


\bibitem[\protect\citeauthoryear{Yang, Dai, Yang, Carbonell, Salakhutdinov, and
  Le}{Yang et~al\mbox{.}}{2019}]%
        {yang2019xlnet}
\bibfield{author}{\bibinfo{person}{Zhilin Yang}, \bibinfo{person}{Zihang Dai},
  \bibinfo{person}{Yiming Yang}, \bibinfo{person}{Jaime Carbonell},
  \bibinfo{person}{Ruslan Salakhutdinov}, {and} \bibinfo{person}{Quoc~V Le}.}
  \bibinfo{year}{2019}\natexlab{}.
\newblock \showarticletitle{Xlnet: Generalized autoregressive pretraining for
  language understanding}.
\newblock \bibinfo{journal}{\emph{arXiv preprint arXiv:1906.08237}}
  (\bibinfo{year}{2019}).
\newblock


\bibitem[\protect\citeauthoryear{Yates, Nogueira, and Lin}{Yates
  et~al\mbox{.}}{2021}]%
        {yates2021pretrained}
\bibfield{author}{\bibinfo{person}{Andrew Yates}, \bibinfo{person}{Rodrigo
  Nogueira}, {and} \bibinfo{person}{Jimmy Lin}.}
  \bibinfo{year}{2021}\natexlab{}.
\newblock \showarticletitle{Pretrained Transformers for Text Ranking: BERT and
  Beyond}. In \bibinfo{booktitle}{\emph{WSDM}}. \bibinfo{pages}{1154--1156}.
\newblock


\bibitem[\protect\citeauthoryear{Yih, Toutanova, Platt, and Meek}{Yih
  et~al\mbox{.}}{2011}]%
        {yih2011learning}
\bibfield{author}{\bibinfo{person}{Wen-tau Yih}, \bibinfo{person}{Kristina
  Toutanova}, \bibinfo{person}{John~C Platt}, {and}
  \bibinfo{person}{Christopher Meek}.} \bibinfo{year}{2011}\natexlab{}.
\newblock \showarticletitle{Learning discriminative projections for text
  similarity measures}. In \bibinfo{booktitle}{\emph{CoNLL}}.
  \bibinfo{pages}{247--256}.
\newblock


\bibitem[\protect\citeauthoryear{Yin, Hu, Tang, Daly, Zhou, Ouyang, Chen, Kang,
  Deng, Nobata, et~al\mbox{.}}{Yin et~al\mbox{.}}{2016}]%
        {yin2016ranking}
\bibfield{author}{\bibinfo{person}{Dawei Yin}, \bibinfo{person}{Yuening Hu},
  \bibinfo{person}{Jiliang Tang}, \bibinfo{person}{Tim Daly},
  \bibinfo{person}{Mianwei Zhou}, \bibinfo{person}{Hua Ouyang},
  \bibinfo{person}{Jianhui Chen}, \bibinfo{person}{Changsung Kang},
  \bibinfo{person}{Hongbo Deng}, \bibinfo{person}{Chikashi Nobata},
  {et~al\mbox{.}}} \bibinfo{year}{2016}\natexlab{}.
\newblock \showarticletitle{Ranking relevance in yahoo search}. In
  \bibinfo{booktitle}{\emph{KDD}}. \bibinfo{pages}{323--332}.
\newblock


\bibitem[\protect\citeauthoryear{Zhang, Wang, Zhang, Tang, Jiang, Xiao, Yan,
  and Yang}{Zhang et~al\mbox{.}}{2020}]%
        {zhang2020towards}
\bibfield{author}{\bibinfo{person}{Han Zhang}, \bibinfo{person}{Songlin Wang},
  \bibinfo{person}{Kang Zhang}, \bibinfo{person}{Zhiling Tang},
  \bibinfo{person}{Yunjiang Jiang}, \bibinfo{person}{Yun Xiao},
  \bibinfo{person}{Weipeng Yan}, {and} \bibinfo{person}{Wen-Yun Yang}.}
  \bibinfo{year}{2020}\natexlab{}.
\newblock \showarticletitle{Towards Personalized and Semantic Retrieval: An
  End-to-End Solution for E-commerce Search via Embedding Learning}.
\newblock \bibinfo{journal}{\emph{arXiv preprint arXiv:2006.02282}}
  (\bibinfo{year}{2020}).
\newblock


\bibitem[\protect\citeauthoryear{Zhao, Liu, Liu, Tang, Guo, Shi, Wang, Gao, and
  Long}{Zhao et~al\mbox{.}}{2020a}]%
        {zhao2020memory}
\bibfield{author}{\bibinfo{person}{Xiangyu Zhao}, \bibinfo{person}{Haochen
  Liu}, \bibinfo{person}{Hui Liu}, \bibinfo{person}{Jiliang Tang},
  \bibinfo{person}{Weiwei Guo}, \bibinfo{person}{Jun Shi},
  \bibinfo{person}{Sida Wang}, \bibinfo{person}{Huiji Gao}, {and}
  \bibinfo{person}{Bo Long}.} \bibinfo{year}{2020}\natexlab{a}.
\newblock \showarticletitle{Memory-efficient Embedding for Recommendations}.
\newblock \bibinfo{journal}{\emph{arXiv preprint arXiv:2006.14827}}
  (\bibinfo{year}{2020}).
\newblock


\bibitem[\protect\citeauthoryear{Zhao, Wang, Chen, Zheng, Liu, and Tang}{Zhao
  et~al\mbox{.}}{2020b}]%
        {zhao2020autoemb}
\bibfield{author}{\bibinfo{person}{Xiangyu Zhao}, \bibinfo{person}{Chong Wang},
  \bibinfo{person}{Ming Chen}, \bibinfo{person}{Xudong Zheng},
  \bibinfo{person}{Xiaobing Liu}, {and} \bibinfo{person}{Jiliang Tang}.}
  \bibinfo{year}{2020}\natexlab{b}.
\newblock \showarticletitle{Autoemb: Automated embedding dimensionality search
  in streaming recommendations}.
\newblock \bibinfo{journal}{\emph{arXiv preprint arXiv:2002.11252}}
  (\bibinfo{year}{2020}).
\newblock


\bibitem[\protect\citeauthoryear{Zou, Xia, Ding, Song, Liu, and Yin}{Zou
  et~al\mbox{.}}{2019}]%
        {zou2019reinforcement}
\bibfield{author}{\bibinfo{person}{Lixin Zou}, \bibinfo{person}{Long Xia},
  \bibinfo{person}{Zhuoye Ding}, \bibinfo{person}{Jiaxing Song},
  \bibinfo{person}{Weidong Liu}, {and} \bibinfo{person}{Dawei Yin}.}
  \bibinfo{year}{2019}\natexlab{}.
\newblock \showarticletitle{Reinforcement learning to optimize long-term user
  engagement in recommender systems}. In \bibinfo{booktitle}{\emph{KDD}}.
  \bibinfo{pages}{2810--2818}.
\newblock


\bibitem[\protect\citeauthoryear{Zou, Xia, Du, Zhang, Bai, Liu, Nie, and
  Yin}{Zou et~al\mbox{.}}{2020a}]%
        {zou2020pseudo}
\bibfield{author}{\bibinfo{person}{Lixin Zou}, \bibinfo{person}{Long Xia},
  \bibinfo{person}{Pan Du}, \bibinfo{person}{Zhuo Zhang}, \bibinfo{person}{Ting
  Bai}, \bibinfo{person}{Weidong Liu}, \bibinfo{person}{Jian-Yun Nie}, {and}
  \bibinfo{person}{Dawei Yin}.} \bibinfo{year}{2020}\natexlab{a}.
\newblock \showarticletitle{Pseudo Dyna-Q: A reinforcement learning framework
  for interactive recommendation}. In \bibinfo{booktitle}{\emph{WSDM}}.
  \bibinfo{pages}{816--824}.
\newblock


\bibitem[\protect\citeauthoryear{Zou, Xia, Gu, Zhao, Liu, Huang, and Yin}{Zou
  et~al\mbox{.}}{2020b}]%
        {zou2020neural}
\bibfield{author}{\bibinfo{person}{Lixin Zou}, \bibinfo{person}{Long Xia},
  \bibinfo{person}{Yulong Gu}, \bibinfo{person}{Xiangyu Zhao},
  \bibinfo{person}{Weidong Liu}, \bibinfo{person}{Jimmy~Xiangji Huang}, {and}
  \bibinfo{person}{Dawei Yin}.} \bibinfo{year}{2020}\natexlab{b}.
\newblock \showarticletitle{Neural Interactive Collaborative Filtering}. In
  \bibinfo{booktitle}{\emph{SIGIR}}. \bibinfo{pages}{749--758}.
\newblock


\bibitem[\protect\citeauthoryear{Zou, Zhang, Cai, Ma, Cheng, Shi, Wang, Cheng,
  and Yin}{Zou et~al\mbox{.}}{2021}]%
        {zou20201pretrained}
\bibfield{author}{\bibinfo{person}{Lixin Zou}, \bibinfo{person}{Shengqiang
  Zhang}, \bibinfo{person}{Hengyi Cai}, \bibinfo{person}{Dehong Ma},
  \bibinfo{person}{Suqi Cheng}, \bibinfo{person}{Daiting Shi},
  \bibinfo{person}{Shuaiqiang Wang}, \bibinfo{person}{Zhicong Cheng}, {and}
  \bibinfo{person}{Dawei Yin}.} \bibinfo{year}{2021}\natexlab{}.
\newblock \showarticletitle{Pre-trained Language Model based Ranking in Baidu
  Search}. In \bibinfo{booktitle}{\emph{KDD}}.
\newblock


\end{thebibliography}

\clearpage

\section{Appendix}
\subsection{Implementation Details}
\noindent\textbf{Computation resources}. We implement the proposed retrieval model with PaddlePaddle (version 1.6.2) \footnote{https://github.com/PaddlePaddle/Paddle}, an open-source library for developing and deploying deep learning models. The model is pretrained and finetuned with 32 and 8 NVIDIA V100 GPUs (32GB Memory), respectively, on our distributed training platform.

\noindent\textbf{Parameter settings}.
We use 6-layer Transformers as the query and document encoders. In the input layer, we tokenize a given query or document title into Chinese characters as integer tokens, and map them into a set of embedding vectors with size 768. For each Transformer layer, we set the dimension of each hidden token representation as 768,  the number of heads as 12, i.e., each head produces a 64 dimensional output. Besides, we set the number of context codes (i.e., $m$) as 16, and the dimension of the compression layers as 256.
For model optimization, we use an Adam optimizer, and set the learning rate as 2e-5 and the batch size as 160 in all stages. For each training stage, we apply 4,000 warm up steps for the learning rate, and a 0.01 decay rate afterwards. During the training, we use a 0.1 dropout rate for all the layers to random drop the attention weights. Other unmentioned details are set as the same as vanilla ERNIE model.

\begin{figure}[!t]
    \centering
    \includegraphics[width=0.45\textwidth]{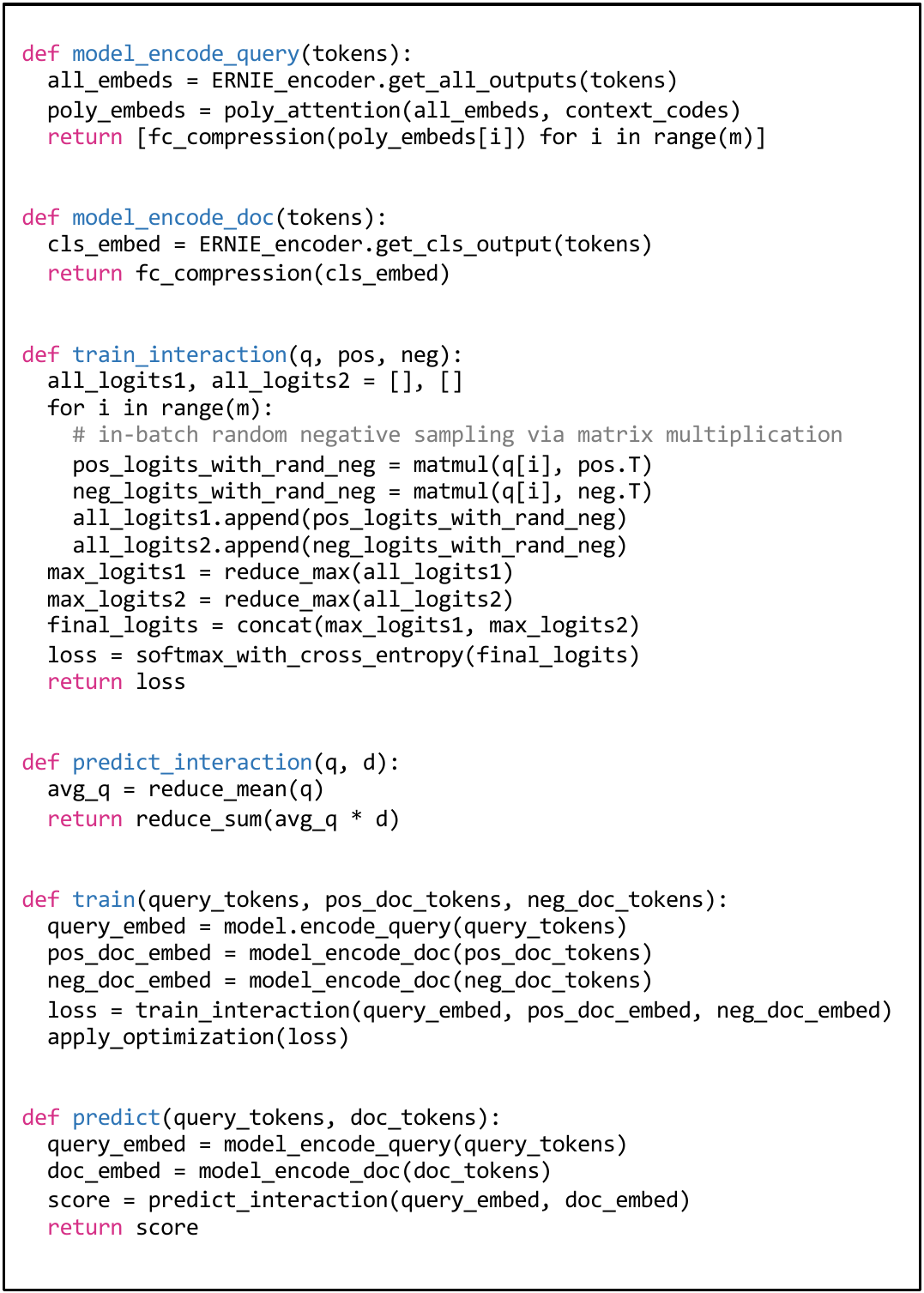}
    \caption{Pseudo code of the model training and prediction.}
    \label{fig:code}
\end{figure}

\noindent\textbf{Implementations}.
We present the pseudo code to depict the overall implementation of our method for the sake of reproducibility. Note that the the in-batch random negative samples are implemented in \textsf{train\_interaction()} by matrix multiplications, which is very efficient in practice. Such implementation is also adopted by other related work (e.g., https://github.com/chijames/Poly-Encoder).

\subsection{Details of Embedding Quantization}
In particular, the output query and document embeddings are all quantized from float32 to uint8. According to output embeddings on a large-scale validation dataset, we calculate the data range ($s_i^{min}$, $s_i^{max}$) for each dimension $i$ of the output embeddings. Then, we divide the data range into $L=255$ equal intervals of Length $Q_i$, where  $Q_i=(s_i^{max}-s_i^{min})/L$. For the value $r_i$ in dimension $i$ of a given output embedding, when performing quantization, its quantized index $QI_i(r_i)$ is calculated by
\begin{equation*}
   QI_i(r_i)=\lfloor(r_i - s_i^{min})/Q_i \rfloor.
\end{equation*}
This index is in range [0, 255], which can be representation as a 8-bit integer. When performing online inference, we recover a quantized value $\tilde{r_i}$ by
\begin{equation*}
\tilde{r_i} = QI_i(r_i) * Q_i + Q_i/2 + s_i^{min}
\end{equation*} to approximate $r_i$. This quantization only causes a small loss of precision which is ignorable for inference, but achieves significant improvements on development efficiency. The quantitative experimental results can be found in Table \ref{tb:ablation}. On the document side, this local quantization helps the system further save huge storage cost. On the query side, it significantly reduces the transmission cost, as the relevance computation for a given query would be distributed to multiple workers. Through the local quantization, the transmission and storage overheads further decrease to 1/4.

\begin{table}[t]
\caption{Ablation study on the improvements of our model.} \label{tb:ablation}
\scalebox{+0.93}{
\begin{tabular}{l|l|c|c|c}
\hline \hline
ID & Model                  & Search log & Manual data & Storage / doc \\ \hline\hline
0  & Base                   & 2.15       & 1.76        & 3072 bytes          \\ \hline
1  & 0 + Poly               & 2.23       & 1.81        & 3072 bytes          \\ \hline
2  & 1 + IBN & 2.33       & 2.06        & 3072 bytes          \\ \hline
3  & 2 + Compression        & 2.26       & 2.00        & 1024 bytes          \\ \hline
4  & 3 + Quantization       & 2.27       & 2.01        & 256 bytes           \\ \hline\hline
\end{tabular}}
\end{table}

\begin{table}[t]
\caption{PNR values when varying the scoring method.} \label{tb:diff_predict}
\scalebox{+0.95}{
\begin{tabular}{l|c|c}
\hline \hline
Method & Search log & Manual data \\ \hline\hline
$\mathrm{max}_{i=1}^{m} P_i \cdot C'$  & 2.205        & 1.955           \\ \hline
$\left( \frac{1}{m}\sum_{i=1}^{m} P_i \right) \cdot C'$  
 & 2.205        & 1.987      \\ \hline
$P_0 \cdot C'$ & 2.161        & 1.940          \\ \hline
$P_1 \cdot C'$ & 2.176        & 1.969         \\ \hline
$P_2 \cdot C'$ & 1.131        & 1.909        \\ \hline
$P_3 \cdot C'$ & 2.204        & 1.976        \\ \hline\hline
\end{tabular}}
\end{table}

\subsection{Offline Ablation Study}
We further conduct several offline experiments to study the impact of some details of the retrieval model. For the experiments, we use two validation datasets, i.e., a search log dataset and a manually-labeled dataset. The construction of the two types of datasets can be found in Section \ref{sec:data_mining}. The manually-labeled dataset is the same as we introduced in Section \ref{sec:datasets}, and the search log data is isolated from the large-scale training data, which contains 300,000 query-document pairs. Note that all the following experiments are conducted for the model trained after intermediate finetuning (i.e., Stage 3).

\subsubsection{Comparison with vanilla bi-encoder.}
Compared with vanilla ERNIE-based bi-encoder, our retrieval model is quite different, i.e., facilitated with poly attention mechanism, in-batch negative sampling strategy, and compression \& quantization. To clarify the influence of each of these differences, Table \ref{tb:ablation} shows the offline experiments of how these distinct features are layered up in our model. Note that the base (i.e., vanilla ERNIE-based bi-encoder) is optimized with hinge loss, where the margin is set to 0.1. We can see from the table that 1) the poly-attention and in-batch negatives (denoted as IBN) can significantly improve the model performance on both datasets, 2) the compression would largely reduce the storage cost, but slightly sacrifice the effectiveness, 3) the quantization is shown to be loss-free w.r.t. the PNR metric on the offline datasets, which is very promising to be adopted online.

\subsubsection{The training-prediction inconsistency.}
As mentioned in Section~\ref{sec:model_arch}, we apply inconsistent schemes to finalize the output score during training and prediction. Here, we conduct experiments to investigate how the inconsistency would affect the performance of the model. Table~\ref{tb:diff_predict} shows the PNR values of the model using different scoring methods on three extract validation datasets. Note that the model used here is an intermediate version, and thus the results might not be aligned with other experiments. In the table, the first row represents the scoring method used in training (i.e., Eq. (\ref{eq:poly_score_train})), the second row represents the scoring method used in prediction (i.e., Eq. (\ref{eq:poly_score_pred})), and $P_0$ to $P_4$ indicate using fixed single global representation for the prediction, who are randomly sampled from the all 16 of them. We can see that our adopted mean-pooling method (i.e., the second row) performs similarly to the scoring method used in training. Thus, such inconsistency does not undermine the performance of the model during prediction.

\end{document}